\documentclass[acmssmall]{acmart}
\usepackage{array}
\AtBeginDocument{%
  \providecommand\BibTeX{{%
    \normalfont B\kern-0.5em{\scshape i\kern-0.25em b}\kern-0.8em\TeX}}}

\setcopyright{acmcopyright}
\copyrightyear{2024}
\acmYear{2024}
\acmDOI{}

\acmConference[Computers as Bad Social Actors]{Computers as Bad Social Actors: Dark Patterns and Anti-Patterns in Interface that Act Socially}{2024}{}
\acmPrice{}
\acmISBN{}

\begin{document}

\title[Computers as Bad Social Actors]{Computers as Bad Social Actors: Dark Patterns and Anti-Patterns in Interfaces that Act Socially}


\author{Lize Alberts}
\email{lize.alberts@cs.ox.ac.uk}
\affiliation{%
  \institution{University of Oxford}
  \city{Oxford}
  \country{United Kingdom}
}
\affiliation{%
  \institution{Stellenbosch University}
  \city{Stellenbosch}
  \country{South Africa}
}

\author{Ulrik Lyngs}
\email{ulrik.lyngs@cs.ox.ac.uk}
\affiliation{%
  \institution{University of Oxford}
  \city{Oxford}
  \country{United Kingdom}
}

\author{Max Van Kleek}
\email{max.van.kleek@cs.ox.ac.uk}
\affiliation{%
  \institution{University of Oxford}
  \city{Oxford}
  \country{United Kingdom}
}




\begin{abstract}

Technologies increasingly mimic human-like social behaviours. Beyond prototypical conversational agents like chatbots, this also applies to basic automated systems like app notifications or self-checkout machines that address or `talk to' users in everyday situations. Whilst early evidence suggests social cues may enhance user experience, we lack a good understanding of when, and why, their use may be inappropriate. Building on a survey of English-speaking smartphone users (n=80), we conducted experience sampling, interview, and workshop studies (n=11) to elicit people's attitudes and preferences regarding \textit{how} automated systems talk to them. We thematically analysed examples of phrasings/conduct participants disliked, the reasons they gave, and what they would prefer instead. One category of inappropriate behaviour we identified regards the use of social cues as tools for manipulation. We describe four unwanted tactics interfaces use: agents playing on users' emotions (e.g., guilt-tripping or coaxing them), being pushy, `mothering' users, or being passive-aggressive. Another category regards pragmatics: personal or situational factors that can make a seemingly friendly or helpful utterance come across as rude, tactless, or invasive. These include failing to account for relevant contextual particulars (e.g., embarrassing users in public); expressing obviously false personalised care; or treating a user in ways that they find inappropriate for the system's role or the nature of their relationship. We discuss these behaviours in terms of an emerging `social' class of dark and anti-patterns. Drawing from participant recommendations, we offer suggestions for improving how interfaces treat people in interactions, including broader normative reflections on treating users respectfully.

\end{abstract}


\begin{teaserfigure}
  \includegraphics[width=\textwidth]{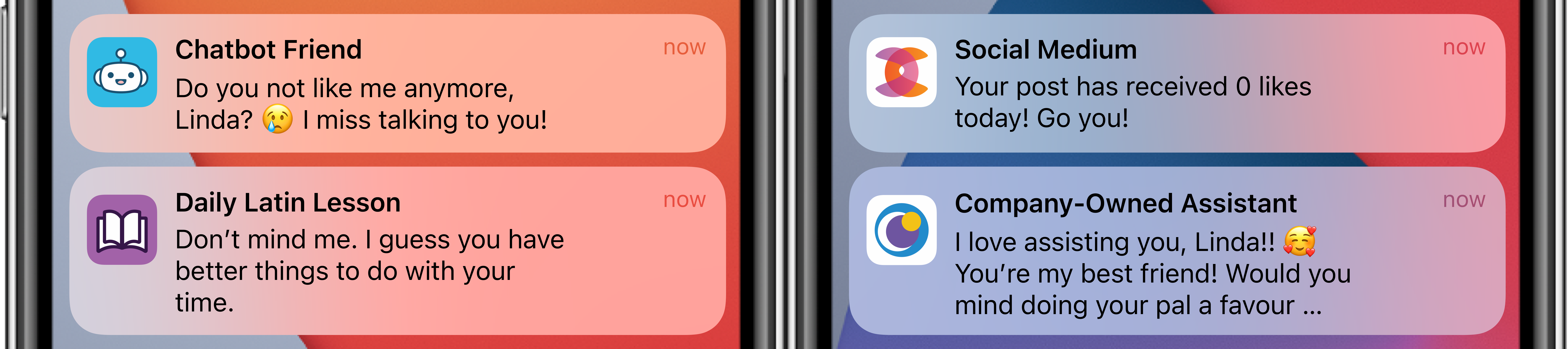}
  \Description[Examples of app notifications using manipulative, rude, or tactless language.]{Four notifications on an iPhone lock screen. Notification 1: "Chatbot Friend: Do you not like me anymore, Linda? :( I miss talking to you!". Notification 2: "Daily Latin Lesson: Don't mind me. I guess you have better things to do with your time.". Notification 3: "Social Medium: Your post has received 0 likes today! Go you!". Notification 4: "Company-Owned Assistant: I love assisting you, Linda!! <3 You're my best friend! Would you mind doing your pal a favour..." }
  \caption{Satirical examples of smartphone app notifications behaving as social actors.}
  \label{fig:CABSA notification examples}
\end{teaserfigure}


%
\maketitle

\section{Introduction}

Automated systems increasingly address or talk to people as if they were thinking/feeling agents. This ranges from sophisticated (digital, virtual, or embodied) conversational agents (CAs) that maintain open-ended dialogues, to basic automated systems like app notifications or self-checkout machines that frame messages as if they were a person directly addressing the user. Whereas prior research has highlighted some benefits of socially conforming/anthropomorphic interfaces, such as fostering a more intuitive and engaging user experience \cite{nass1994computers, abdi2018scoping} and increasing user trust \cite{seeger2017we}, we do not yet have a good understanding of when, and why, the use of different social cues may be inappropriate. Beyond emerging concerns regarding agents based on large language models (LLMs) generating biased, toxic or misleading language \cite{Renee, bender, taxonomy}, there is a lack of empirical data on how pragmatic factors, like the broader social context and perceived goal of automated social actions, affect how such behaviours are received. 

In this paper, we explore the misuse and abuse of social cues in interfaces, how users prefer to be spoken to, and what it may mean for automated systems to treat them more appropriately and respectfully in interaction. For this, we look beyond the more prototypical conversational user interfaces (CUIs) like chatbots, voice-assistants, and social robots, to any interface that behaves as a social actor:  using social cues (e.g., affective expressions, first-person pronouns, etc.), or merely `saying things’, be it in writing, voice, or gesture, as if addressing the user. We use the term \textit{social interfaces}\footnote{This is not to be confused with the use of `social interface' in development sociology as ``a critical point of intersection between different lifeworlds, social fields or levels of social organization'' \cite[p.243]{long2003development}. Here, `social' describes the behaviour of the interface itself.}  to refer to this broader class. 

Our first concern is the use of social interface elements to manipulate. When interfaces mimic patterns in natural human-human social interaction, they encourage users to respond in social or emotional ways---as if it were a person talking to them---by tapping into known social/anthropomorphic biases \cite{reeves1996media,nass2000machines,nass1994computers}. In doing so, behavioural designers may unduly influence people’s behaviour, in ways that bypass their rational (reflective) agency, into doing things they may have otherwise been reluctant to do---not unlike the infamous \textit{dark} \textit{patterns} in interaction design \cite{mathur2019dark, narayanan2020dark, mathur21,lukoff2021can}.\footnote{Broadly, these are tactics that apply insights from behavioural psychology to steer users' behaviour towards promoting another stakeholder's interests.} So far, most of the dark patterns literature has focused on interface elements (e.g., on shopping websites or cookie consent banners) that subtly `trick’, seduce, or pressure users into behaviours they may have otherwise been reluctant to, like spending money or compromising their privacy.

Such socially manipulative tactics have already started emerging---most commonly in smartphone notifications. Smartphone apps increasingly frame their notification messages in conversational (person- or even friend-like) ways to encourage users to stick to their goals, complete tasks, buy products, or otherwise increase app engagement. Some even pressure users by eliciting guilt or empathy, using emotive language like ``I miss you’’, or seeming annoyed at users for ``ignoring'' them. Whilst the use of social cues to influence user behaviour may not be necessarily \textit{bad}, they can certainly be exploited. We characterise the inappropriate use of such tactics as a social class of dark patterns: where designers try to manipulate people towards certain outcomes by encouraging them to treat an interface as a thinking/feeling agent (whose feelings and judgment they should care about), rather than an inanimate platform. 

Another concern we investigate, is the risk of social interfaces failing to exhibit appropriate nuance for a given social situation (which is notoriously difficult\footnote{See Suchman's discussion of interaction as the ``contingent coproduction of a shared sociomaterial world'' \cite[p.23]{suchman2007human}.}) when they proactively `speak to’ people in everyday settings.  If messages are made to seem as if they are coming from an agent specifically addressing the user, they may appear invasive, insensitive, or even offensive in the wrong context---especially when they are scripted, repetitive or generic. Following the common distinction between dark and anti-patterns \cite{macdonald2019anti,gray2018dark}---i.e., interface elements that fail to work as intended---we characterise \textit{social} anti-patterns as automated social behaviour that bother users for reasons that the designers failed to anticipate: particularly for seeming, in some sense, socially inappropriate. 

Driven by these concerns, we used qualitative methods to explore the following research questions with end-users: 

\begin{itemize}
\item \textbf{RQ1}: When do automated social behaviours seem manipulative or pressuring, and why?
\item \textbf{RQ2:} When do automated social behaviours seem otherwise off-putting or inappropriate, and why?
\item \textbf{RQ3:} How do people prefer automated systems ``speak'' to them, generally?
\item \textbf{RQ4:} What contextual factors affect their preferences?
\end{itemize}

Our study consisted of four complementary phases \cite{lukoff2021design,tran2019modeling}. The first was an exploratory survey (n=80) with adult smartphone users, prompting them to describe occasions when they disliked how automated systems spoke to them. This was followed by three studies, using a subset of participants (n=11) from the first: (1) a week-long experience sampling study, capturing \textit{in-situ} examples of notification messages that spoke to people in ways they disliked, annotated with details of what about them was undesirable, and why; (2) individual semi-structured interviews, allowing participants to elaborate on nuances in their preferences, and (3) a group-based exploratory workshop (n=3-4) with activities exploring variation in preferences between application domains. This combination of methods was designed to help us triangulate and refine our findings: whereas the first phase allowed us to gain an initial understanding of stand-out negative experiences across social interfaces, the latter three helped us gain a better understanding of contextual, individual and domain variability. 

Using an integrated, reflective approach to analyse our data into codes and themes \cite{o2021mixing,braun2019reflecting}, we generated three sets of themes to gain a better understanding of what makes the use of social cues more or less inappropriate. The first set concerns the manipulative use of social cues, as dark patterns (RQ1). The second set regards counterproductive uses of social cues, constituting social anti-patterns (RQ2). The final set regards participant suggestions for improved social acting (RQ3, RQ4), where we describe themes generated from participant suggestions regarding how they would prefer, and believe they deserve, to be treated by interfaces that ``talk'' to/at them.

Our primary contributions can be summarised as follows: 
\begin{enumerate}
\item We contribute to the literature on nudging/dark patterns by identifying and characterising an emerging \textit{social} class of dark patterns.
\item We use our findings to critically engage with research in the Computers Are Social Actors paradigm by proposing when, and why, the use of social cues in interfaces may be counterproductive---constituting `social' anti-patterns.
\item Based on our analysis of end-user preferences and suggestions, we offer recommendations to help prevent undesirable forms of automated social acting.
\item We integrate normative considerations regarding how users feel they \textit{deserve} to be treated by automated systems as specific duties for designers: laying the foundations for a framework unpacking what it means for interfaces to treat people respectfully in interaction.
\end{enumerate}

Our findings highlight important factors that can affect the perceived appropriateness of automated social behaviours. The first is the use of social/emotionally manipulative tactics: from our participants’ examples and experiences, we identify four tactics that already appear in interface design. These are \textit{agents playing on emotions} (e.g., guilt-tripping, coaxing, or eliciting empathy),\textit{ agents being pushy} (e.g., dictating, pressuring or nagging), \textit{agents mothering} (e.g., behaving like a concerned parent), and \textit{agents being passive-aggressive} (e.g., conveying dissatisfaction or judgement)---all of which undermine users’ sense of autonomy. We also identify a range of factors that can make seemingly innocuous or beneficent social behaviours inappropriate. This includes embarrassing users by addressing them in public spaces, or recommendations that, for contextual reasons, seem like they are mocking users' weaknesses or insecurities. We also found that users can find certain social pleasantries off-putting for seeming insincere, especially when incentives are obviously self-interested or messages generic. Moreover, using casual, friend-like language can likewise be off-putting when it feels misaligned with the agent's perceived role or relationship with the user. 

Our work contributes to the literature on users’ experiences of CUIs, particularly regarding expectation violations \cite{hadi2019humanizing,lucas2017role,meng2021emotional}, as well as emerging research in Human-AI Interaction \cite{amershi2019guidelines,li2022assessing,berk}. It also contributes to recent work anticipating harms related to LLM-based dialogue agents \cite{Renee,bender,taxonomy}, which are expected to play increasingly active social roles in people's daily lives. However, ours is the first paper to treat in a unified manner the expanding set of systems that ``speak'' to and address users. Whilst prior work has mainly focused on the more overtly human-like versions of these, such as chatbots or social robots, our treatment admits a much larger set of interfaces that, while perhaps less socially capable, nonetheless create social situations through their communicative actions and use of social protocol. In doing so, we argue that they introduce kinds of risks that are best understood in social terms: as social actors failing to meet what we would expect from intentional agents in similar social situations. 

With the conceptual and empirical contributions of this paper, we hope to inspire a focused discussion on how automated systems treat people in interactions, and how it can be improved.

\section{Background and related work}
\subsection{When is social acting (in)appropriate?} 

The Computers Are Social Actors (CASA) paradigm is centred on understanding people's anthropomorphic tendencies when interacting with computers. Starting with a series of simple lab studies in the 90s, CASA became a research paradigm centred on the finding that people apply social norms and expectations (or \textit{scripts} \cite{reeves1996media}) to their interactions with technology, even without being prompted by obvious anthropomorphic elements \cite{nass2000machines,nass1994computers}.\footnote{CASA is a specific application of the \textit{Media Equation} \cite{reeves1996media}, the idea that people behave as if they treat ``mediated'' life as real life; interacting with communication technologies in fundamentally social ways. This may include treating computers as if they have folk-psychological states (e.g., beliefs, intentions, or desires), or reacting to moving pictures on a screen as if the events were taking place in real life \cite{nass2000machines}.} According to Reeves and Nass \cite{reeves1996media}, these responses occur even when people do not have corresponding beliefs that justify their behaviours, or offer rational arguments for why it would be silly to---we seem to apply such scripts \textit{mindlessly}. 

Complementing CASA, findings in neuroscience suggest that cues to human-like animacy can strengthen our anthropomorphic tendencies, as they engage brain networks associated with social cognition. These can be bottom-up (e.g., how an entity looks/behaves), as well as top-down (e.g., beliefs of human-like commonalities) \cite{cross2016shaping}. In interface design, common examples are \textit{identity cues} (e.g., giving a virtual agent a name or face), \textit{non-verbal cues} (e.g., affective speech, gendered voices or pausing as if thinking), and \textit{verbal cues} (e.g., human-like mannerisms in responses) \cite{grimes2021mental}.\footnote{Other (top-down) cues include personified descriptions of AI systems: beyond the countless AI personifications that populate the media (including the term \textit{AI} itself), CUIs like social robots are often explicitly marketed as ``friends'' and ``your next family member'' that ``can’t wait to meet you'' \cite[p.60]{donath2020ethical}.} Such attributes help to make the system appear more like an intentional `agent' speaking to the user, rather than a medium for people to communicate with each other \cite{gambino2020building}, strengthening social/emotional responses. 

However, heightening expectations of a system's social capabilities, and then falling short of them, may severely harm users' overall experience and satisfaction \cite{6,33}. Moreover, researchers have started identifying individual factors (e.g., users' cognitive styles \cite{lee2010more}, personality traits \cite{kocielnik2021can}, age \cite{folstad2020users,laitinen2016social}) and contextual factors (e.g., the domain of application \cite{svenningsson2019artificial}) that may affect users' evaluation of social interfaces,\footnote{For example, in a focus group exploring older adults' perspectives on social robot carers/companions, Laitinen \textit{et al.} \cite{laitinen2016social} found that several participants expressed concern that toy-like robots may be perceived as patronising/infantilising, a form of stigmatising that older adults already struggle with.}  raising the question of what social cues are desirable, and in which contexts. Using certain social cues in the wrong situations may even have backfiring effects, e.g., making angry customers angrier \cite{hadi2019humanizing,lucas2017role}, or stressed users more stressed \cite{meng2021emotional}. In an analysis of 500 customer reviews of chatbot apps, Svikhnuskina \textit{et al.} \cite{svikhnushina2020social} found that, whilst current chatbots offer certain benefits to users, they fail to meet various contextual expectations, including repeating responses, going off-topic, and being perceived as ``rude''. Moreover, they found frequent mentions of chatbots being ``unnaturally supportive'' to the point of discomfort. 

Whilst studies like these point to factors that may render social cues ineffective, undesirable and even offensive in certain contexts, studies of users’ experience with social interfaces are generally limited in terms of sampling and ecological validity (e.g., using lab-based/wizard-of-oz techniques). User experience (UX) studies with conversational systems also tend to focus on the usability of specific systems (and often focus on basic metrics like `enjoyment' or `satisfaction' \cite{abdi2018scoping}). As such, we aimed to delve deeper into users' more general experience with the range of social interfaces they encounter in their daily lives, even ones that are not typically considered `conversational'. 

Beyond social acting generally, we wanted to investigate the use of social cues in behavioural design. To lay the groundwork for this, the next subsection briefly introduces the concepts of nudging and dark patterns, and distinguishes anti-patterns.

\subsection{Distinguishing nudges, dark and anti-patterns}
Behavioural design involves applying knowledge of how certain (psychological, social, material) factors influence human behaviour to interface design. In the case of nudges and dark patterns, this involves knowledge of people's general cognitive heuristics and biases,\footnote{That is, systematic deviations from rational judgment in human behaviour \cite{amos}.}, which designers may either try to counter (by prompting careful reflection) or harness (utilising biases strategically). Both typically involve manipulating choice architectures---how options are framed or presented---such that certain options are more likely to be chosen. Their key difference lies therein that nudges are typically aimed at promoting the interests of the user, whereas dark patterns are typically aimed at promoting those of the provider; although, as we will show, this boundary can be fuzzy. As such, we briefly discuss how we distinguish our understanding of dark patterns.

Originally, \textit{nudging} was proposed as an unobtrusive and ethical form of paternalism, ``without forbidding any options or significantly changing their economic incentives'' \cite[p.6]{nudge}. This idea has been applied broadly in HCI, from preventing unwanted mistakes to helping users stick to their goals (see Caraban's \cite{caraban201923} review). Over time, however, the term has been used to describe a wider range of strategies for `beneficent' behaviour manipulation, including more or less overt/intrusive, and sometimes controversial means \cite{brabazon2015digital,dodsworth2021state}. This can involve inducing fear, using deceptive tactics, or invoking a sense of shame or social pressure towards some desired end \cite{caraban201923,dolan2010mindspace}.

Dark (design) patterns,\footnote{These are also sometimes called \textit{deceptive patterns, }to avoid racial connotations. However, as the tactics we consider do not necessarily involve deception, we find this term misleading.} on the other hand, may harness various nudging (or even coercive) tactics to get users to take certain actions that they may otherwise have actively avoided \cite{lukoff2021can,mathur2019dark}. This typically involves inducing some misleading mental model by modifying choice architectures---\textit{manipulating the decision space} (e.g., by placing unequal burdens on choices or strategically omitting options) and/or the \textit{flow of information} (e.g., through deceptive framing or hiding important information) \cite{mathur21}---such that users are led to make certain inferences and take certain actions. 

'Dark pattern' has been used to describe multiple related (but not collectively required) aspects of objectionable interfaces \cite{mathur21}. Whilst some researchers define it in terms of the \textit{intention} to exploit users towards self-interested goals \cite{conti2010malicious,gray2018dark}—contrary to `beneficent' nudging—others find it objectionable for reasons beyond intention. According to a recent review by Mathur \textit{et al.} \cite{mathur21}, this includes facts about the {\itshape interface} (e.g. being deceptive, coercive, or manipulative), the \textit{mechanisms} of influence (e.g., subverting user intent or preferences), and the \textit{effects} of the interface design (e.g., benefiting other stakeholders and/or harming users).

Given the breadth of these definitions, it is not always clear when a nudge counts as a dark pattern, as even well-intentioned forms of behavioural design (e.g., reminders to eat healthily) can meet some of these criteria (e.g., harming users by shaming them, or undermining their sense of autonomy). Moreover, intentions are rarely clear-cut,\footnote{A behaviour that serves the provider's interests can still be seen as, in some regards, serving the user.} and what is in someone's `best interest' is often a matter of framing. Given these ambiguities, we use the term `dark patterns' in a broad sense, regardless of intention, to refer to design patterns that involve inducing certain emotions (e.g., guilt, fear, shame) or misleading mental models (e.g., a misconception of the system or their actions) to manipulate user behaviour towards specific ends---involving mechanisms or goals that the user is unaware of or did not consent to.\footnote{We add the latter, as we would not consider mechanisms for behaviour change that people purposefully use to help them keep to their goals as `dark’, but we may consider it a dark pattern if a mental health app encourages engagement in unpleasant (e.g., fear-inducing or autonomy undermining) ways as an added feature by default.}

Most of the literature on dark patterns has focused on how information is phrased or presented in graphic user interfaces (GUIs), most prominently cookie consent banners \cite{7,9,281}, ads on online platforms \cite{soc,pol}, and e-commerce websites \cite{mathur2019dark,com,comm}. Recently, a handful of papers have started considering dark patterns in the ways that other modalities interact with people, e.g.,social robots \cite{cute,shamsudhin2022social}, auditory interactions with voice user interfaces (VUIs) \cite{owens2022exploring}, and proxemic interactions with movable platforms \cite{proxemic}. In section three, we extend this by characterising a \textit{social} class of dark pattern that can be implemented, in some form, in everything from GUIs, to VUIs, tovirtual agents, to robots.

Whereas dark patterns may involve inducing frustrating, or upsetting experiences in strategic ways, other design patterns may produce negative consequences purely because of oversight or bad design, known as anti-patterns. 

\subsubsection{What are anti-patterns?}
The term `anti-pattern’ was introduced by Brown \textit{et al.} to describe ``a commonly occurring solution to a problem that generates decidedly negative consequences'' \cite[p.7]{brown1998antipatterns}. Rather than strategically manipulating design elements to bring about certain outcomes, here negative consequences result \textit{unexpectedly} due to oversight on the designer's part (e.g., making the wrong button too easy to press, or the right button too difficult). Possible causes are a lack of knowledge, insufficient experience in solving a particular type of problem (e.g., implementing a pattern/theoretical approach poorly), or applying a useful or trendy pattern in an inappropriate context \cite{bolanos2011antipatterns}. Another common cause is designers failing to carefully consider aspects like different users’ experience or their contextual needs and goals \cite{bolanos2011antipatterns} (e.g., image recognition software that only recognises certain skin tones, or text that colourblind users find unreadable). 

Having laid the theoretical foundations and reviewed related work, the next section introduces `social' dark and anti-patterns as a vocabulary to identify a particular class of risks for social interfaces.

\section{Characterising `social' forms of dark and anti-patterns}
In our literature review, we found two papers relating to social cues being used manipulatively by interface designers \cite{cute,shamsudhin2022social}. Both of these limit their focus to social robots, and neither empirically engage with user experience. 

Lacey and Caudwell argue that the `cute’ aesthetic features of home robots should be considered a dark pattern in that they elicit a ``powerful affective bond'' from users, masking their potentially harmful (i.e., privacy-imposing) features \cite[p.374]{cute}. Moreover, by leading the user to assume the role of ``caregiver'', they suggest that the robot's childlike cuteness gives the user a false sense of authority that may obscure the powers of influence the technology has over them \cite[p.378]{cute}. This is echoed by Shamsudhin and Jotterand \cite{shamsudhin2022social}, who frame social robot design as an ``inherently persuasive project'', as users are led to ``believing, at least temporarily, that the robot is human-like, has life-like properties, can be trusted, and there is value in the creation and maintenance of this human–robot relationship'' \cite[p.95]{shamsudhin2022social}. However, a (cute) appearance is only one of many social cues in robots that can elicit social/emotional responses:  
\begin{quote}
Robots that incorporate social cues such as gaze, proximity, and facial expressions, push our Darwinian buttons ... and effectively coerce us into interacting with them socially \cite[p.1915]{springer}.
\end{quote}
Any combination of verbal or nonverbal cues could be used in manipulative ways \cite{shamsudhin2022social}; for instance, suggesting disappointment or anger with body language or facial expressions to elicit guilt or shame. Text or voice alone can use sentiment or tone to convey judgment or emotion strategically (e.g., ``It makes me sad that you would not…'', or ``It would make me happy if you would…'')---especially when coupled with expressions of affection or familiarity, like calling the user their ``best friend'' or addressing them by name. Whilst in some contexts, such phrases may be appropriate (e.g., in robot/chatbot companions or toys), they may be more objectionable in others (e.g., when trying to get users to do or consent to something they are reluctant to). Such social design patterns can be implemented in any range of interfaces, not just sophisticated CAs like social robots. However, risks may be amplified when systems behave in more convincingly human-like ways, or if a user has built a trusting `relationship' with a specific system-as-agent over time \cite{shamsudhin2022social}. These are merely a few examples of how interface designers may use social cues to manipulate, in more or less exploitative ways. 

Following our earlier definition, we characterise \textit{social dark patterns} as the use of \textit{social }design patterns (i.e., social cues or interaction patterns) for inducing certain emotions or misleading mental models (e.g., a misconception of the system’s capacities) to manipulate user behaviour towards certain ends; particularly, involving mechanisms or goals they are not aware of or did not consent to. As in dark patterns generally, this involves utilising/exploiting knowledge of specific cognitive biases in people's behaviour: in this case, social/anthropomorphic biases that lead people to treat an interface as a thinking/feeling agent, whose feelings and judgment they care about, rather than a platform \cite{nass2000machines}.  

Apart from unduly influencing people's behaviour, social behaviours may also be inappropriate for contextual reasons, such as exhibiting a lack of tact or situational sensitivity (although these are not mutually exclusive). We characterise \textit{social anti-patterns} as social design patterns that are applied poorly or in an inappropriate context, such that they negatively impact user experience. For instance, a designer could think users may enjoy it when a system addresses them by a pet name (e.g., \textit{sweetie}), but certain users may find this off-putting, or even offensive (e.g., being deemed patronising or sexist). Such risks are especially high in less sophisticated forms of automated social acting, like generic messages that are made to seem like it is personalised to the user, or addressing them directly, at that moment, when it is not.

We introduce this vocabulary to talk about particular risks that come with raising expectations of interfaces as social actors. With our conceptual and empirical contributions, we hope to encourage further research on risks of this nature: particularly, how interfaces treat people in social interaction.

\section{Methods}

\begin{figure}[h]
  \centering
  \includegraphics[width=\linewidth]{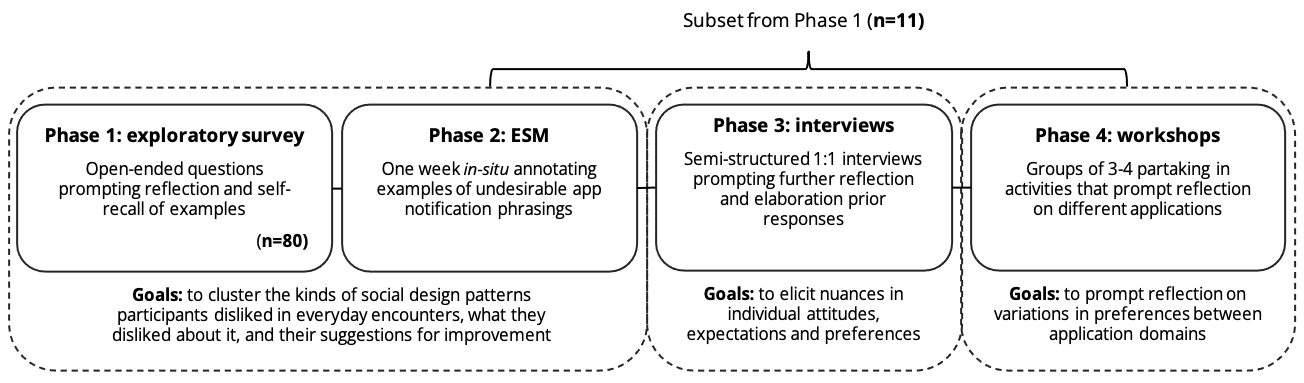}
  \caption{Flowchart of study design and methods}
  \Description[A diagrammatic representation of the methods used in this study]{A flowchart of boxes filled with text, each representing a different phase of the study: ``Phase 1: exploratory survey, Open-ended questions prompting reflection and self-recall of examples, Jisc Online surveys (n=80)'' followed by ``Phase 2: ESM, One week in-situ capturing and reflecting on examples of undesirable notification messages from apps on their phone'', followed by ``Phase 3: interviews, Semi-structured 1:1 interviews prompting further reflection and elaboration on survey and ESM responses'', followed by ``Phase 4: workshop Groups of 3-4 partaking in activities that prompt reflection on different application domains of social cues in interfaces''. There is a bracket indicating that Phases 2-4 used a ``subset of participants from phase 1: (n=11)''.  Underneath Phase 1 and 2 (combined), there is the following text: ``Goals: to find patterns in the kinds of social design patterns participants dislike in everyday encounters, what they disliked about it, and their suggestions for improvement''. Underneath Phase 3: ``Goals: to elicit nuances in individual attitudes, expectations and preferences'', and underneath Phase 4: ``Goals: to prompt reflection on variations in preferences between application domains, and how participants believe users deserve to be treated in each''}
  \label{Diagram 1}
\end{figure}

Our study had four phases: an exploratory survey (n=80), followed by a week-long study using an experience sampling method (ESM) \cite{van2017experience}, individual semi-structured interviews, and a group-based exploratory workshop. The latter three phases used a subset of participants from Phase 1 (n=11). A summary of the phases is shown in Figure \ref{Diagram 1}.  All interviews and workshops took place online. 

Using a mixed qualitative methods \cite{o2021mixing,tran2019modeling,lukoff2021design} (or \textit{intra-paradigm} \cite{o2015advanced}) design, we integrated qualitative data from the four phases to triangulate our findings. This combination of methods was designed to counteract the limitations of each phase, and help us to uncover nuances in our participants' experiences and preferences. Data was thematically analysed in an iterative, integrated way using a combination of Braun and Clarke's \cite{clarke2015thematic,braun2019reflecting} reflexive approach and O'Reilly \textit{et al.}'s \cite{o2021mixing} integrated analytic approach, described below.

Our aim was exploratory: to start identifying socially manipulative tactics that are used in current interfaces, as well as everyday situations in which people can find certain automated social behaviours otherwise inappropriate. 

This study was approved by our university's Central University Research Ethics Committee (CUREC). Data, coding, and study materials are available on the Open Science Framework (OSF).\footnote{OSF link will be made available with the final published version.}

\subsection{Phase 1: Exploratory survey of 80 smartphone users} 

The first phase was an online survey consisting mainly of free-text questions. The aim was to collect examples of social behaviours that respondents found bothersome or off-putting in automated systems, and to start gaining an understanding of what they disliked about those encounters. 

\subsubsection{Recruitment}
Recruitment of participants was done through Prolific Academic, as well as a Twitter post using our university's departmental research group account. We used the following inclusion criteria:

\begin{itemize}
\item Being from/primarily residing in a predominantly English-speaking country
\item Being 18 years or older
\item Owning a smartphone
\end{itemize}

We focused this initial exploration on adults, as we suspected younger participants may experience social interfaces differently.\footnote{For instance, children may be less likely to find it inappropriate for interfaces to speak to them in playful or paternalistic ways, or ``play pretend''.} We chose the second criteria as the researchers spoke English as a common language. We intended to recruit all participants via Twitter, but included Prolific when we had a smaller turnout than expected.\footnote{Prolific survey respondents were compensated at a rate of £10 an hour, with an estimated completion time of 21 minutes (£3.50 base pay), although most took less than 10 minutes. Twitter respondents completed the survey voluntarily, but all were compensated equally for partaking in the further phases.} We then recruited around 20 participants at a time and analysed the data until we started seeing saturation in our findings. 

A total of 84 participants filled in the Phase 1 survey: 19 recruited via Twitter, and 65 via Prolific. Of our Twitter responses, we excluded three for not meeting the location criteria, and one for not consenting to our data usage terms, leaving 80 participants in the final data set. 53\% identified as male, 46\% as female, and 1\% preferred to not disclose. Over half (54\%) were based in the UK, and 63\% were between the ages of 21 and 40. Table \ref{tab:1} summarises the demographic information. As the sorts of social interfaces people commonly encountered at the time were limited,\footnote{This study was conducted in August, 2022.} we found 80 participants sufficient. This relatively low number was also more feasible for doing thematic analysis across the qualitative data set.

\begin{center}
\begin{table}
\caption{Demographics for Study 1.}
\label{tab:1}
\begin{tabular}{ | c | c |} 
  \hline
  \textbf{Country of residence} & UK (54\%), Canada (24\%), Ireland (14\%), South-Africa (6\%), US (3\%)\\  
  \hline
 \textbf{Age range} & 21-30 (35\%), 31-40 (28\%), 41-50 (14\%), 51-60 (13\%), 18-20 (8\%),  61-70 (4\%)\\
  \hline
 \textbf{Gender} & Male (53\%), Female (46\%), Prefer not to disclose (1\%)\\ 
  \hline
\end{tabular}
\end{table}
\end{center}

\subsubsection{Procedure}
To administer the survey, we used \textit{Jisc online surveys},\footnote{\url{https://www.onlinesurveys.ac.uk/}} a platform that had data protection agreements in place with our institution. We designed the survey to capture a broad collection of examples, and explanations, of encounters with interfaces `speaking' to participants in ways they found bothersome or off-putting. 
We were also interested in their general attitudes to different forms of treatment by automated systems: for instance, using more or less formal tones; expressing more or less emotion; or treating them more as a friend or a customer.
Our aim was primarily to \textit{understand reasons} for negative experiences, rather than making inferences about the \textit{relative frequency} of negative experiences. As such, the questions were designed for qualitative analysis: capturing users' examples/experiences of undesirable automated social behaviours and their reasons, in their own words. 

As the most common type of social interface, the survey first asked about negative encounters with smartphone notifications: whether participants had ever been ``\textit{particularly annoyed or put off by the tone or phrasing of an app notification}''. 
If so, they were asked to list any examples they could recall, followed by an explanation of what it was about it that they disliked. 
We then asked the same question regarding social interfaces more broadly: whether participants had been ``\textit{annoyed or put off by other automated systems (recordings, chatbots, self-checkout machines, etc.) `speaking' to you as if they were a real person}''.  

To understand their general preferences, respondents were asked to rate agreement on a five-point Likert scale with the following statements: 

\textit{I like it when an automated system...}
\begin{itemize}
    \item\textit{ ...talks to me in a friendly/chatty tone (e.g. greets me, addresses me by name, uses conversational language)
\item ...speaks like it has thoughts or feelings of its own (e.g. "I think/feel...")
\item ...has other human-like qualities (e.g. uses a human-sounding voice, has a name, has a face)
\item ...talks to me as if we're friends}
\end{itemize}

This was followed by a question to ``\textit{elaborate a bit on the above (e.g. when you do or don't, and why)}''. Given our qualitative focus, our analysis focused on the patterns in respondents' elaboration on the extent to which they liked different manners of speaking, rather than on the Likert scores \textit{per se}: their purpose was primarily to prompt reflection. 

Towards the end, the survey asked participants if they had ``\textit{any further thoughts to share on the topic of an automated system or artificial agent acting as if it were a person}''. The survey concluded by asking respondents if they were interested in participating in an upcoming study on the same topic, which we used for recruitment for the other study phases.




\subsection{Phase 2: Experience  sampling method}
Whereas the survey allowed us to rapidly capture data (from a broader sample of participants) regarding stand-out negative encounters with social interfaces, the second phase used an ESM to capture \textit{in-situ} examples of undesirable social behaviours as they happened, annotated with participant's descriptions of what bothered them, factors that contributed to their experiences and suggestions for improvement. This gave us access to relevant contextual information about particular experiences that the survey was less likely to capture, and circumvented the survey's reliance on respondents' ability to recall experiences from memory. For this, we focused on app notifications as a basic form of social interface with which most smartphone users would be familiar (and likely encounter multiple times daily).

\subsubsection{Recruitment} For the next set of studies (Phases 2-4), we invited 24 participants who completed the previous study and expressed interest in participating again, using no further exclusion criteria. 16 agreed to participate but five dropped out before starting, leaving 11 (demographics in Table \ref{tab:2}). 
\begin{table}
    \centering
\caption{Demographics for Phases 2-4.}
\label{tab:2}
    \begin{tabular}{|c|c|c|c|c|} \hline 
         \textbf{Participant}&  \textbf{Country of residence}&  \textbf{Age range}& \textbf{Gender} &\textbf{Workshop group}\\ \hline 
         P1&  UK&  31-40& Female &G2\\ \hline 
         P2&  Ireland&  31-40& Female &G1\\ \hline 
         P3&  Canada&  21-30& Female &G3\\ \hline 
         P4&  UK&  51-60& Male &G1\\ \hline 
         P5&  Canada&  31-40& Female &G2\\ \hline 
         P6&  UK&  21-30& Male &G3\\ \hline 
         P7&  South-Africa&  21-30& Male &G2\\ \hline 
         P8&  South-Africa&  21-30& Male &G1\\\hline 
 P9& UK& 41-50&Female &G2\\\hline 
 P11& Ireland& 41-50&Male &G3\\\hline
 P12& UK& 31-40&Male &G3\\\hline 
    \end{tabular}
    
\end{table}

\subsubsection{Procedure}
At the start of Phase 2, participants were emailed a consent form and information sheet detailing the goals, requirements and data processing procedures for Phase 2-4. We informed participants that the first study required enabling all app notifications on their smartphones (with the exception of personal messaging apps like WhatsApp, or email, where they could keep their current notification settings). For one week (at least five and at most seven days), participants were asked to capture and annotate all the notification phrasings that bothered them, as they occurred.  The information sheet, and prior survey, informed them that we were mainly interested in \textit{``the way in which such systems `speak' to them''}, rather than other aspects of the notifications. 

Each participant was given access to a set of editable Microsoft Word tables (one for each day), located in a private folder in the first author's university Nexus 365 OneDrive SharePoint account. They were asked to add entries whenever they encountered an undesirable notification (or take screenshots during the day and add them by the end, which we instructed them how to do). Table \ref{Table 3} shows the column headings of the entry form. Participants were offered five spaces for daily entries but were instructed on how to add more if needed.

\begin{center}
\begin{table}
\caption{\label{Table 3}Headings in the daily diary tables for the ESM.}
\begin{tabular}{ | m{3cm} | m{1.5cm}| m{3cm} | m{3cm}| m{3cm} |}
  \hline
    \textbf{Quote} (or image/ description/ paraphrase) \textbf{of the notification message} & \textbf{App that sent it} (optional) & \textbf{What was it about this example that bothered you?} & \textbf{Please describe any contextual factors} (e.g., your current mood, task, environment, or personal background) \textbf{that you think may have contributed to your bad experience} & \textbf{If you could change the notification in any way, how would you improve it?}\\  
  \hline
\end{tabular}
\end{table}
\end{center} 

\subsection{Phase 3: Semi-structured interviews}
As the format of Phases 1 and 2 (boxes to fill out) may have incentivised overly concise or reductive responses, this phase offered participants the opportunity to elaborate on their responses and related questions.

\subsubsection{Procedure}
After the ESM, we conducted 1:1 semi-structured interviews (15-30 min) with each participant over Microsoft Teams. We adapted the questions to each participant's survey responses and ESM entries, asking them to elaborate and reflect deeper on the causes of their dissatisfaction with specific notifications, as well as nuances or exceptions to their preferences. For instance, if a participant stated in the survey that they dislike when a system talks to them as a person, we asked them if they could think of any context where they might feel differently. We also tried to unpack the source of any dissatisfaction with interfaces behaving in social ways: whether they believed it was down to a lack of technical sophistication, or something more fundamental. If participants were bothered by, e.g., obviously generic messages, we asked about the extent to which they would prefer if technologies used personal data to tailor more to them.

\subsection{Phase 4: Exploratory workshops}
The aim of the final phase, exploratory workshops (45 min), was to prompt group discussion and reflection on how their preferences may vary between domains and applications. Another goal was to elicit more information about how participants felt they \textit{deserved} to be treated in different domains, exploring what appropriate or respectful treatment might mean to them. 

\subsubsection{Procedure} 
We invited all participants from the previous two phases to one of three online workshops conducted using Microsoft Teams and the online whiteboard tool.\footnote{Miro https://miro.com} Through a series of four Miro activities (mainly involving generating sticky notes) workshop participants (n=3-4) speculated together on their preferences regarding different social interfaces. First, they were told to select two interfaces out of a choice of six (7 min): \textit{an app that motivates you to stick to your goals}, \textit{a virtual personal receptionist}, \textit{a smart home assistant }, \textit{a tutoring chatbot for small children}, \textit{a robot carer/companion for the elderly}, or \textit{a chatbot therapist app} . We then asked them to generate sticky notes on (1) ``desirable ways for  the system to treat/talk to people in each context (i.e., in a way that is respectful/appropriate)'', and (2) ``undesirable (off-putting, disrespectful, unethical ways for an automated system to treat people in each context'' (10-15 min). They were then asked to reflect on specific rights they believe users should have for how they are treated in each domain (15 min). Finally, they were given the opportunity to briefly reflect on how their preferences may differ between other contexts and to cluster similar preferences together (5 min). Here, we gave them all the initial options, as well as eight other applications (e.g., a talking fridge, an automated vehicle, etc.) as examples. After describing each task, the researcher left the call to allow participants to discuss it amongst themselves.

\subsection{Data and analysis}
\subsubsection{Data preparation}

At the end of Phase 2, we collated each participant's (5-7) daily forms into a single PDF document for analysis. During Phase 3, we took audio and video recordings of each (15-20 min) interview via Microsoft Teams and stored them in a secure institutional Nexus 365 OneDrive SharePoint folder. The lead author then downloaded the auto-generated transcripts from each interview (done by Microsoft Teams) and, following the recordings, edited them by hand for correctness and de-personalisation. The transcripts were then thematically analysed. Finally, in Phase 4, we took audio and video recordings of each (45 min) workshop, and downloaded the completed anonymous Miro worksheets (containing sticky notes filled with text) as PDFs. We used the recordings of group discussions to help us analyse the worksheets more accurately. 

\subsubsection{Analytic approach}
Rather than treating each phase separately, as in similar HCI studies \cite{tran2019modeling,lukoff2021design}, our data was analysed using O'Reilly \textit{et al.}'s \cite{o2021mixing} integrated analytic approach: using a mixture of inductive and deductive enquiry to iteratively refine themes and codes generated from multiple qualitative data sets \cite{o2021mixing}. We chose this approach as our methods were designed to complement and enhance each other, offering different perspectives and nuances on users' experiences with social interfaces. Not only did this enable a deeper and richer engagement with our issues of interest \cite{o2021mixing}, but also helped us to find patterns of inconsistencies and similarities between contexts through critical comparison (similar to grounded theory approaches \cite{chun2019grounded}). 

To code and generate themes from our data, we used Braun and Clarke's \cite{clarke2015thematic,braun2019reflecting} reflexive approach. In what follows, we highlight a few assumptions underlying our analysis \cite{o2021mixing}. As our study focused on a largely unexplored area of dark patterns, our analysis was neither completely inductive (data-driven) or deductive (theory-driven). On the one hand, we were, to some extent, driven by theory, as our particular understanding of dark patterns drove the kinds of research questions we posed and helped us to recognise similar tactics. However, rather than using a pre-specified codebook, we expected the types of dark patterns that would emerge in a social context to differ from those found in other kinds of interfaces, and so we were committed to understanding what \textit{users} found manipulative or otherwise off-putting in these kinds of interfaces and why. We also did not want to assume that a design pattern is `dark' purely because it seems to fit our theoretic criteria, and so we actively encouraged our participants repeatedly reflect on the extent to which different patterns may be positive in different contexts. 

\subsubsection{Analysis process}

We started our analysis by open-coding the survey results from Phase 1, guided by our research questions. In particular, we distinguished examples of social acting that contained social design patterns that seemed manipulative (i.e., framing things in ways that could be construed as pressuring users to take certain actions) (RQ1) and reasons that users found certain social design patterns otherwise off-putting (RQ2, RQ4). We also started coding patterns in suggestions from participants on how to improve features they disliked (RQ3). However, we also inductively coded patterns that went beyond our research questions. In particular, we noticed patterns in how participants feel they deserve to be treated on more principled grounds, which we decided to incorporate into the remaining study designs. By the end of this process, we started combining codes into low-level themes.

During Phase 2, our codes and themes from Phase 1 helped us to deductively code similar extracts, but were refined or updated when patterns of nuances or exceptions were found. The ESM helped us collect more examples of manipulative tactics, richer data on contextual reasons why users found social design patterns off-putting, as well as more specific suggestions for improvement. As a part of this process, we iterated back over our prior codes and started clustering them into higher-level themes.

The codes and themes generated thus far helped us to plan the interview questions that would be most useful to ask in Phase 3, so as to further test and refine our findings. This phase followed a similar process of iteration, abstraction, and critical comparison. 

Finally, the Phase 4 workshop data were coded separately as they were most usefully clustered by application domain, since each workshop focused on two particular social interfaces. As the workshop was more future-facing (anticipating differences in preferences between hypothetical social interfaces), rather than looking for things participants disliked and would like to improve, we looked for social design patterns that groups decided were desirable (RQ3) and undesirable/inappropriate (RQ2, RQ4) relative to each domain. Following our inductive coding of more normative/principled considerations for how users feel they should be treated, we asked groups to reflect on this for each application domain and thematically coded this as well. We used these results on domain variability to expand/add nuance to our prior findings during the write-up stage. 

By the end of this process, our codes and themes were ordered within three sets: \textit{social dark patterns}, \textit{social anti-patterns}, and \textit{suggestions for improved social acting}. Thematic coding was conducted using NVivo 1.7.1. 

\section{Results}

\subsection{Data collected}

In total, survey participants contributed 84 examples of notifications that bothered them (median number of examples per participant = 1, min = 0, max = 3) and 62 examples of bothersome social behaviours by other automated systems (median = 1, min = 0, max = 3). In Phase 2, the 11 participants collectively contributed 125 diary entries about notifications they found bothersome. Two people participated for five days, one for only four (due to special circumstances), and the rest for all seven. Of these, only one participant experienced no bothersome notifications (stating this in their form for every day). Seven participants had entries of bothersome notifications on all of the days they participated, two on 71\% of their days, and one on 51\%. The median number of entries per participant was 8 (min = 0, max = 24).

In Phase 4, we conducted two workshops with four participants (G2, G3), and one with three (G1). The application domains chosen by the groups to discuss were: a smart home assistant (G3), a tutoring chatbot for small children (G2), a robot carer/companion for the elderly (G1), and a chatbot therapist app (G1, G2, G3). 

The next section summarises the themes and sub-themes we arrived at by the end of our integrated analysis process.

\subsection{Social dark patterns: manipulative uses of social cues (RQ1)}

Under this theme, we coded excerpts relating to social behaviours our participants found ``manipulative or pressuring'' (RQ1). Our participants' examples and descriptions showed clear overlap with known dark pattern tactics (e.g., being dishonest, pressuring, inducing guilt/shame), but were typically framed as the (social) behaviour of a digital \textit{agent}, rather than the design of a platform (although all participants knew the messages were scripted by a person). One of our participants explained, ``\textit{I anthropomorphise more than I would like. When a machine talks to me, I can't help but think of it as a person and how I'm relating to it, and treating it, as a person.}'' (P50, \textbf{S}\footnote{We will mark the data source with the following acronyms: \textbf{S} = survey (Phase 1), \textbf{E} = experience sampling (Phase 2), \textbf{I} = interview (Phase 3), \textbf{W} = workshop (Phase 4). For data from Phase 1-3, quotes are accompanied with an anonymous participant ID, e.g. ``P50''; for data from Phase 4, quotes accompanied with the number of the workshop group where the data was generated e.g., on a Miro worksheet.}). We constructed four themes capturing types of tactics our participants described: ``agents playing on emotions'', ``agents being pushy'', ``agents mothering', and ``agents being passive-aggressive''. 

\subsubsection{Agents playing on emotions} 
The first theme (for which contributing codes applied to 21\% of survey respondents, 10\% of all ESM entries, and 5/11 interviews) was the use of emotional manipulation by systems that behave as social actors (e.g., using first-person pronouns or addressing the user directly) and express certain emotions (especially disappointment or sadness) in order to get users to \textit{do} something. This either involved appealing to their empathy (making the user feel bad for the `agent')  or appealing to their self-image (e.g., coaxing them or making them feel bad about themselves). Describing their general experiences with bothersome app notifications, one participant wrote: ``\textit{Some [say] they're sorry to see me go or that they will miss me, which is false. Some try to make me feel bad for ignoring them}'' (P32, \textbf{S}). Another described how the system tries to steer them against their own desires, illustrating the manipulative aspect: ``\textit{Feeling sad when I `hurt' it, or worse, ignore it. But I WANT to be able to ignore my phone and my stupid apps.}'' (P50, \textbf{S}). Even if participants did not find this sort of tactic persuasive, there was general consensus that it should not be used as they found it annoying and/or unethical. Several participants criticised such tactics for being dishonest about the system's capacities, expressing emotions that were ``false'' or a ``lie'': ``\textit{We all know it is not a person with emotions, so whatever human-like qualities are used, must be a kind of manipulation to get or keep you interested}'' (P13, \textbf{S}). 

\subsubsection{Agents being pushy}
This theme contained references (26\% \textbf{S}; 20\% \textbf{E}; 1/11 \textbf{I}) to an automated system using dictating language (e.g., giving orders), aggression (e.g., shouting, using all caps or exclamation marks) or repetitive nagging to convey a sense of urgency to pressure a user to do a task. These kinds of tactics were found both in self-interested (e.g., marketing) messages and ``beneficent' messages meant to help the user (e.g., reminders or instructions). For instance, one participant disliked the urgent tone of a notification from the popular language learning app Duolingo: ``\textit{Hi, it's Duo. Reminding you to practice French. Got 3 minutes now?}'', explaining that ``\textit{the sense of obligation that is created is unpleasant}'' (P12, \textbf{E}). Several survey respondents also expressed annoyance at self-checkout machines for being ``pushy'':
\begin{quote}
\textit{She comes across as very aggressive and loud. It feels like she's shouting at you. And she's very pushy and impatient. For example, when completing your purchase, she immediately and continuously reminds you to remove all scanned items from the bagging area.} (P5, \textbf{S})
\end{quote}
Whilst such prompts are meant to be helpful, a rapid succession of orders felt to some participants like it was pressuring them to do the task faster, along with inducing a sense of being watched or judged by the `agent'. 

\subsubsection{Agents mothering} 
This theme contained references (8\% \textbf{S}; 5\% \textbf{E}; 6/11 \textbf{I}) to tactics that felt paternalistic or overly helpful in a way that undermined participants' sense of autonomy: ``\textit{It's kind of like a parent you can't chat back to}'' (P8, \textbf{S)}. We chose the term `mothering' based on one of our participant responses: ``\textit{Apps regularly make me feel like they are trying to mother me}'' (P34, \textbf{S}), because it was often the \textit{tone} of the message (e.g., overly helpful or concerned), that made participants feel inappropriately controlled. In a rather poignant example, one respondent complained about their phone's bedtime reminder stating that their bedtime is approaching and that they should wind down soon: ``\textit{I feel like it is too dictating and even though I feel like I am not particularly sleepy or outside having fun I feel bad not to be ready to go to bed at that moment}'' (P70, \textbf{S}). Whilst, in some cases, these are similar to some `pushy' tactics described above, we distinguished this theme for being specifically described in overly protective or parent-like terms. During interview discussions, several participants highlighted the importance of respecting user autonomy, even if they want to act outside of their best interests: ``\textit{you have to empower the person to make that decision. You can't make too many assumptions about their well-being on their behalf.}'' (P12, \textbf{I}). However, there were also examples where this tactic was used for more self-interested means: ``\textit{Stressed? (With tear face, worried-looking emoji). Why not take a break and play?}'' (P5, \textbf{E}), in the case of a game app prompting a user to open it.

 \subsubsection{Agents being passive-aggressive} 
Rather than being overtly pushy or dictating, tactics under this theme regarded automated systems using more covert means to convey dissatisfaction or judgment through tone or implication (e.g., sarcasm) as a person would (6\% \textbf{S}; 0\% \textbf{E}; 0/11 \textbf{I}). A few of the excerpts referenced Duolingo here, e.g., ``\textit{Hey it's Lily, Duo says you're ignoring him so he's sent me}'' (P13, \textbf{S}). However, most mentions of  ``passive-aggressive'' behaviour did not include concrete examples. 

\subsection{Social anti-patterns: counterproductive uses of social cues (RQ2)}
As opposed to the behavioural tactics above that strategically induce certain emotions or mental models to manipulate, another set of themes (guided by RQ2) related to social design patterns that upset users for reasons that the designers failed to anticipate. At a high level, we distinguished between the themes \textit{contextually insensitive or tactless}, \textit{inappropriate use of tone} (e.g., talking to users in inappropriately friend-like, parental, or childlike ways), and \textit{obviously faking capacities} (i.e., systems pretending to be aware/sincere when it is clear they are not). 

\subsubsection{Contextually insensitive or tactless}
The first theme contained references (50\% \textbf{S}; 30\% \textbf{E}; 10/11 \textbf{I} ) to automated systems interrupting users or saying something (seemingly innocuous) at an inappropriate time, such that it seems insensitive or offensive by implication. For example, one participant mentioned self-checkout machines telling them to ```\textit{remove the item from the area' ... sometimes it's a false flag and the system doesn't realize but keeps repeating as if accusing you of stealing something}'' (P65, \textbf{S}). Several excerpts mentioned systems being oblivious to contextual factors that make a suggestion irrelevant (such as suggesting they start something they are already doing), or even hurtful. For instance, one ESM participant was bothered by a notification telling them ``you've achieved 73\% of your step goal'', explaining: ``\textit{I went mountain climbing. First time in my life and I felt proud of myself}'' (P8, \textbf{E}). Another participant received a notification for ordering alcoholic drinks at 3 p.m., which they found insensitive given that they ``\textit{used to drink a bit more than I like to}'' (P3, \textbf{E}).


\subsubsection{Inappropriate use of tone}
This theme contained references (35\% \textbf{S}; 22\% \textbf{E}; 8/11 \textbf{I}) to tones or \textit{ways of speaking} that participants found off-putting for not being aligned with the social role of, or nature of their relationship with, the system-as-agent. The first was an inappropriate use of a friend-like tone, conveying a sense of unwarranted closeness (e.g., addressing the user by name or pet names, or using affectionate emojis). This often made participants feel uncomfortable, especially in marketing or professional service contexts. In such contexts, perceived incentives seemed to taint how agreeable and friendly behaviours seemed: ``\textit{...what are the incentives? Why is this thing doing this? What is it trying to accomplish? Is it actually trying to help me? Is it trying to help the company?}'' (P7, \textbf{I}). Some participants also described it as ``crossing a boundary'' as they felt that is something that needs to be ``earned'': ``\textit{It's a term of endearment that you get over time, you know}'' (P5, \textbf{I}). To explain the strangeness, two participants compared such behaviours to a random person \textit{``coming up to them in the street''} or \textit{``in the shop''}, saying it would be strange even if a person spoke in such a level of familiarity ``out of nowhere'': ``\textit{There is definitely a line of friendliness that is sometimes crossed and it usually comes across as fake/try-hard/creepy.}'' (P7, \textbf{S}).  

There was also a dislike among our participants of interfaces treating them as if they were a child (e.g. pitying them, praising them, or explaining more than necessary). Examples include being put off by ``\textit{an investing app telling me `well done' and `good job' etc when I was using it}'' (P42, \textbf{S}) or a self-checkout machine ``\textit{saying obvious things like `don't forget your receipt!}''' (P28, \textbf{S}). Multiple participants commented that they find such behaviours ``patronising''. This also included frustrations around being treated as if one needs special treatment, e.g., ``\textit{[the self-checkout machine's] loudness makes me think that she thinks I have a hearing problem or something}" (P5, \textbf{S}). Such patterns often overlapped with \textit{mothering} tactics described above, which seemed to put participants off regardless of manipulative elements. Again, this bore on a lack of appropriateness given the nature of their relationship with the agent, as some participants explained it can be acceptable when someone you are ``close to'' or ``care about'' speaks to you in a certain way, but not an interface: ``\textit{It's so different, it doesn't know me, like, `Who are you to have that right?'}'' (P8, \textbf{I}). 
 
Rather than sounding condescending, another common complaint was social interfaces using child or teen-like language (e.g., emojis, \textit{netspeak} or other infantile behaviour---particularly in the context of marketing). Such behaviours came up a few times in the smartphone notifications that participants captured in the ESM (12\% \textbf{E}), and were described as ``try-hard'', ``childish'', and ``annoying''. As one participant elaborated in the interview, ``\textit{I think basic pleasantries like, I don’t know, just like, `please' whatever is fine, but when it’s trying to, like, be your friend or trying to be, like, too relatable ... that type of thing is just kind of cringy}.'' (P3, \textbf{I})

\subsubsection{Obviously faking capacities}
A bothersome behaviour that was mentioned in all phases of study (39\% \textbf{S}; 15\% \textbf{E}; 5/11 \textbf{I}, as well as 3/3 of the workshop groups) was when an interface clearly ``fakes'' capacities for comprehension, care, or concern, typically when it is noticeably a generic or pre-programmed message. A common complaint was that it comes across as ``condescending'' when participants are expected to be easy to fool, or willing to ``play make-believe''. As one participant insisted, ``\textit{I don't want to play make-believe with something ... we know you're robot!}'' (P1, \textbf{I}). Several participants were also bothered by systems ``acting'' like they care about them in a personalised way, when they clearly do not: ``\textit{I just know it's just like programmed to say, like, `Hi, are you having a good day?' and it's just like, no one's actually there asking me that. That's just everyone who opens this app or whatever is going to see that.}'' (P3, \textbf{I}). 

Contrary to early CASA research that was taken to support the use of baseless flattery and pleasantries in systems \cite{nass2000machines}, some of our participants were put off by the obvious insincerity of such acts: ``\textit{This is coming from nowhere. This is not coming from a person who cares about me and wants me to feel good or whatever, or cares about anything. It doesn't have anything. It's empty, it's a machine.}'' (P12, \textbf{I}). Multiple participants expressed a strong desire for ``machines''/automated systems to just act \textit{like what they are}---without trying to seem human-like in any way. Beyond being misleading, several participants suggested that what counts as appropriate for machines may differ from people:
\begin{quote}
\textit{I generally dislike automated systems posing as real human beings as it decreases their authenticity. If I engage with an actual human, I'm expecting realistic human responses. If my conversation is carried out with an automated service I rather expect them to give me straight answers and simple choices, there is no need to introduce elements like pretending they have feelings as it dehumanises the whole experience even more by introducing fake genuine interest on their part} (P28, \textbf{S})
\end{quote}


\subsection{Participant suggestions for improved social acting (RQ3, RQ4)}
Along with our investigation into dark and anti-patterns in social interfaces, we also wanted to explore with our participants---at each phase of study---how they would improve the design of these interfaces if they could. This set of themes was centred on answering our final two research questions (RQ3, RQ4). Whilst we expect variation in preferences between different demographics, we wanted to highlight the suggestions that we found preliminary consensus on. As the ESM was limited to preferences in the context of smartphone notifications, we used the workshops to elicit preferences across other application domains, which we discuss under the relevant themes.

\subsubsection{Machines should ``stay in their role''}
There was a common desire (44\% \textbf{S}; 1\% \textbf{E}; 9/11 \textbf{I})  among participants for automated systems to ``stay in their role'' in terms of acting like a tool, a service, or a lifeless machine (as opposed to a human): ``\textit{It's a bot, I don't need it to pretend and be human, a friend or anything else other than a bot}'' (P66, \textbf{S}). This includes putting effectivity before ``flourishes'' (e.g., being chatty, personable, or making small-talk), as it detracts from the immediate needs of the user. A few participants said that they would not mind simple conversational pleasantries \textit{in theory}, as long as it does not detract from the actual purpose of the system: ``\textit{The more human-like a system is, the more likely I am to feel like I'm supposed to treat it like a human. This is often at odds with the job of whatever the machine is}'' (P49, \textbf{S}). Relatedly, some participants expressed a desire for automated systems generally ``keeping the relationship professional''.  During the interviews, we asked participants to reflect on the extent to which they might feel differently if systems were more sophisticated in the future. However, a few considered it something they would always feel uncomfortable with: ``\textit{I just don't think computers will ever be so much like a human that I would feel comfortable with them interacting with me like a human.}'' (P5, \textbf{I}). 

During the workshops, we identified some contexts in which preferences may differ. One group considered human-likeness more appropriate in the context of smart home devices or robot companions for the elderly (G2, \textbf{W}). They also suggested that a tutoring bot for young children would warrant more of a "casual", "friendly and fun" (G2, \textbf{W}) tone than a formal/professional one.


\subsubsection{Make the baseline `neutral'}
This theme related to participants' descriptions of what a good standard way of speaking would be for automated systems generally. We coded several excerpts preferring `to-the-point'' and ``neutral'' language (61\% \textbf{S}; 10\% \textbf{E}; 36 \textbf{I}). That is, some ``medium'' between sounding too formal/monotonous and informal/excited: ``\textit{The exaggerated fake happiness in the voices is patronising to me}'' (P70, \textbf{S}), ``\textit{Just cut the overexcitement}'' (P1, \textbf{S}). This also included preferences for speaking in a clear, concise and transparent way: ``\textit{not like, `Ohh, today's gonna be a great day. The weather's shining outside' ... just tell me the weather. We don't need the fluff.}'' (P7, \textbf{I}).

\subsubsection{Anticipate relevant contextual factors}
This theme contained references to specific contextual factors that participants believed may affect how social behaviours are received (48\% \textbf{S}; 26\% \textbf{E}; 9/11 \textbf{I}). One factor was individual differences, in which participants mentioned their level of extroversion, personality, mood, age, current activity, and neurodivergence as possibly contributing to their preferences. For instance, a few participants complained about being spoken to/addressed without their consent, due to being shy or private people: ``\textit{Self-checkout machines volume is way too loud and draws attention. I'm shy in public and it gives me anxiety to use self-checkouts}'' (P19, \textbf{S}). Workshop participants considered some aged-related differences: whilst one group suggested children may not find overly excited/helpful tones as patronising as adults (G2, \textbf{W}), another mentioned that elderly participants may be extra sensitive to feeling patronised/infantilised, as they are commonly subject to demeaning treatment (G1, \textbf{W}). Several ESM participants also mentioned more complex personal situations affecting how notifications are received: ``\textit{I don't have a right to tell my health app like, `No. I'm going through an emotional time' or whatever}'' (P3, \textbf{I}). 

Another factor was the passing of time, which may change the nature of a relationship the user has with an agent (making more familiar tones more appropriate), and, conversely, make the same behaviours less appealing: ``\textit{...its first notifications tend to be more useful. And the more you get them, it's just like, `Oh, screw it, I'm over that.'}'' (P8, \textbf{I}).

\subsubsection{Offer means for customisation}
Multiple participants expressed a need for customisation, to have the means to exert more control over how they are ``spoken to`` (13\% \textbf{S}; 9\% \textbf{E}; 5/11 \textbf{I}). A few comparisons were made to human-human contexts, stating that usually, in interacting with a person, one has the ability to negotiate how one prefers to be treated, whereas automated systems (like notifications) generally take more of a ``take-it-or-leave-it'' approach: ``\textit{[With a friend or a partner] at the very least you would have a conversation and deal with it in some way ... Whereas, with the app, there's no way of, kind of, saying, `No, sorry, please just remind me and that's enough}''' (P12, \textbf{I}). In some cases, offering means for changing how one is addressed can be especially important, as in the case of dead-naming trans people: ``\textit{as a trans person, names [are] a bit fraught, especially when my bank app insists on using my legal name which is not the one I go by in daily life}'' (P13, \textbf{S}). 

\subsubsection{General normative considerations}

Finally, through our iterative coding of the different phases, we started noticing excerpts that took more principled stances on how participants believe users \textit{deserve} to be treated, as basic normative standards for \textit{all} social interfaces. Some of these give further justification for preferences/frustrations raised above. 

One theme was respecting user autonomy (8\% \textbf{S}; 2\% \textbf{E}; 7/11 \textbf{I}, 3/3 \textbf{W}) by not letting them asymmetrically bend to the system's needs. In the case of automated systems giving instructions or reminders, one participant complained ``\textit{there's no interlocutor with whom you can have a conversation to try to come to an accommodation or any form of compromise with}'' (P12, \textbf{I}). Specific examples from the workshops included allowing users to choose not to respond to a therapy bot's questions (G2) or to pause the session (G1). In principle, this means treating people as rational agents, rather than objects to control or manipulate: ``\textit{We need to treat people as agents, not as input and output and... processing machines}'' (P12, \textbf{I}), or ``\textit{I'd resent the app for playing on my feelings and maybe be spiteful, not do something on purpose” }(P7, \textbf{I}).

A related theme was to design machines in ways that make their intentions and capacities transparent, so as not to mislead people, which constituted another theme (40\% \textbf{S}; 14\% \textbf{E}; 11/11 \textbf{I}, 2/3 \textbf{W}). This also relates to the theme of respecting user intelligence (3\% \textbf{S}; 5\% \textbf{E}; 1/11 \textbf{I}, 1/3 \textbf{W}) by not treating them as incapable or incompetent (e.g., by over-explaining or withholding important information). 

Another theme was to not ``pigeonhole'' users (14\% \textbf{S}; 6\% \textbf{E}; 5/11 \textbf{I}, 2/3 \textbf{W}) by assuming too much about who they are or what they like. On the topic of personalising services with user data, one interview participant expressed a need for being treated as dynamic and unpredictable: ``\textit{The whole point of being human is that you can act in unpredictable ways, and you can reinvent yourself all the time. And algorithmic profiling does not allow for that}'' (P12, \textbf{I}). Practically, one workshop group suggested that a therapy chatbot should be able to truly ``listen to'' or accommodate individual concerns, rather than just offering advice (G3).

The final theme was respecting people's sense of personal boundaries (11\% \textbf{S}; 1\% \textbf{E}; 4/11 \textbf{I}, 1/3 \textbf{W}). This included concerns around agents ``knowing too much'' about users in a way that feels ``invasive'' or ``creepy''.  One participant suggested that the \textit{sense} of privacy/secrecy can be at least as important as what the system actually knows: ``\textit{I know that you're listening to me, but don't make it quite so obvious. Like, you know, hide in the bushes over there rather than the bushes directly in front of me. Give me at least some kind of figment of privacy'}' (P2, \textbf{I}). 

\section{Discussion and implications}

In this work, we integrated four qualitative methods to elicit user experiences and preferences regarding how automated systems ‘talk’ to them: from app notifications, to self-checkout machines, to chatbots. Our findings highlight important factors that can affect the extent to which social design patterns  (i.e., social cues or interaction patterns) are deemed appropriate. We distinguish between social \textit{dark patterns}, social design patterns that are used to manipulate user behaviour towards certain ends, and social \textit{anti-patterns}, where social design patterns are applied poorly or in an inappropriate context, such that they bother users for reasons the designers failed to anticipate.

From iteratively analysing and coding participants' survey responses, ESM entries, interviews, and workshop group discussions, we generated four themes that capture types of social dark pattern tactics that already appear in interfaces, most prominently in smartphone notifications. We also constructed three themes that describe situational factors that can make certain social design patterns be received badly, such as seeming insensitive, insincere, or invasive. Finally, we constructed five themes that represent specific suggestions, from our participants, on how to improve the ways that interfaces talk to users, including considerations regarding the sorts of design choices that they find unethical or disrespectful in more principled terms.

As an exploratory study, our aim was not to give a definitive overview of all the tactics that could be used as dark patterns in social interfaces, but to characterise an emerging class of dark and anti-patterns, and to understand user attitudes towards examples that already exist in common automated systems. We also do not mean to posit these as people's feelings towards social interfaces generally. Rather, we wanted to shine a lens on everyday contexts in which automated forms of social acting, particularly when talking to/at users proactively, can be received badly, and to start identifying patterns in common reasons for why that can be. Our ultimate aim is to gain a better understanding of what it may mean for an interface to be a \textit{good} (tactful, respectful, constructive) social actor.

\subsection{The misuse and abuse of `social' design patterns}

To our knowledge, this work is the first that treats in a unified manner the expanding set of systems that behave as social actors---not necessarily for their use of overt social/anthropomorphic cues, but for proactively ``saying things'' (in text, sound, or gesture) as if addressing the user. Thereby, even very basic interactive systems like notifications are able to \textit{create }social situations \cite{suchman2007human}, leading to frustrations when these actions seem socially inappropriate: whether it is for certain facts about the system, such as their role or relationship with the user, or facts about the user and their current situation. Knowing well that a system is merely a platform, our participants still tended to describe their frustrations in terms of human-like attributes (e.g., ``\textit{it}'' or ``\textit{she}'' seeming passive-aggressive, insincere, judgmental, or ``knowing'' too much), as they were judging the \textit{behaviour} of the `agent' as indicative of such attributes. In line with findings in the CASA paradigm, the interface is apparently treated as a social actor, as its actions are nevertheless treated (described/experienced) as if coming from an intentional agent---given what the receiver would expect from an intentional agent in that context. 

We anticipate a risk that such intuitive responses---arising from known social/anthropomorphic biases in people--may be harnessed to manipulate user behaviour, in typical dark pattern fashion \cite{mathur2019dark,mathur21}. That is, by encouraging people to treat the interface as an agent whose feelings or judgment they should care about, as a means of pressuring them to act. Our preliminary findings, mainly in the context of smartphone notifications, indicate that such socially manipulative tactics are already emerging, such as eliciting empathy for the `agent', expressing judgment on its behalf, or acting as a concerned parent that controls the user out of care for them. Whilst the deception risk is relatively low in the context of notifications, we believe it it may increase with more sophisticated forms of social interfaces (e.g., chatbots or robots)---especially when interacting with vulnerable populations like children, or the technologically illiterate. This expands on Lacey and Caudwell's \cite{cute} use of \textit{cuteness} as a dark pattern, by positioning it within a broader class of dark patterns that involve leveraging social cues in agent-like systems, offering a vocabulary for future HCI researchers to help identify/talk about similar tactics.

Our findings also offer more insight into how raising expectations of a system \textit{as a social actor} can backfire. Prior UX research has mainly considered expectation violations for chatbots \cite{hadi2019humanizing,lucas2017role,meng2021emotional}, where certain social design patterns (like using affectionate/friend-like language) may make more sense, as users may have the opportunity to build some form of rapport/relationship with the agent (depending on its role). However, when applying the same design patterns in contexts where such familiar/friend-like terms have not been ``earned'', our participants expressed feeling tricked or even ``dehumanised''---especially when the message is obviously generic.

Following the recent successes/hype surrounding dialogue agents based on LLMs, there has been a spike in interest in developing interfaces that behave as social actors. Whilst researchers have started exploring risks related to such technologies, most of the focus thus far has been on the level of language itself: making outputs as ``helpful, honest, and harmless'' as possible \cite{pan2023automatically}, by, for instance, guarding against biased, misleading, toxic, or inaccurate content \cite{bender,Renee,taxonomy}. However, our findings show that even seemingly innocuous (helpful, friendly, `harmless') social actions can cause offence. This pragmatic dimension of social-interaction harms has not yet been empirically explored, as interfaces that are more commonly considered "conversational", like chatbot apps, social robots, and LLM agents, have so far mainly been engaged \textit{with} by users at times of their choosing, rather than taking a more active social role across different contexts. As this is likely to change in future, we offer preliminary empirical evidence of situational risks that are harder to mitigate on a data/language-level alone. 

Whilst there were areas where we identified relative consensus (e.g., participants preferring more professional, emotionless tones for automated systems), this may be due, in part, to the particular domain of focus (i.e., app notifications), and demographic similarities (e.g., predominantly English-speaking adults). However, the fact that there was such an overwhelming backlash to certain ways that social design patterns have been used, shows that more attention needs to be given to how different people prefer to be spoken to/addressed in different contexts, rather than assuming `the friendlier the better', or that what works well in one domain will work in another. With our findings, we hope to initiate a discussion on risks regarding particularly social-interactional harms (e.g., feeling disrespected, judged, or offended) that increase with interfaces behaving as social actors.

\subsection{Towards a more respectful approach to CASA}
Whilst it may be alluring to apply CASA insights to behavioural design, our findings suggest that people are much more socially discerning than a designer may hope---especially after becoming increasingly accustomed to forms of social acting being used across platforms. Rather than ``mindlessly'' treating a social interface as sentient or finding it agreeable, people may, contrarily, be immediately suspicious \textit{because} they suspect that someone is trying to ``get something'' from them. An important general disanalogy with \textit{people} as social actors, is that we expect people to act in service of their own needs and desires, whereas, with a platform, there is the knowledge that it is an artefact designed to serve external interests. Thus, rather than seeming fun or engaging, friend-like or coaxing behaviours are easily perceived as insincere or patronising. Even when systems are designed to be beneficent---proactively motivating or suggesting/reminding users of activities that serve their own interests---users easily get frustrated if this is done in a way that is undeservedly familiar, parental, or merely for knowing that the system \textit{doesn't actually care about them}. 

More fundamentally, this points to a critique of common assumptions underlying behavioural design, in social interfaces or otherwise. In applying general design patterns to steer user behaviour, expecting them to be ``easy'' to manipulate using predefined strategies, it treats the user not as an intelligent, self-defining agent, but a ``mindless'' collection of biases and predictable responses. Rather, interfaces should treat people respectfully: in ways that show ``a commitment to core values that make someone a person" \cite[p.1]{babushkina2022does}, such as their intelligence, rational agency, and sense of self-worth. Some of the normative considerations our participants raised, start to paint a picture of what this may look like, which we integrate and summarise below.

\subsubsection{Be transparent about intentions and capacities}
Interfaces should be transparent, not only about their machinelike nature, but about their actual capacities and the intentions of their stakeholders. Even if participants are not effectively deceived, the very act of trying to deceive them can feel patronising (i.e., as if undermining their intelligence).

\subsubsection{Allow people to negotiate how they are treated}
Rather than treating users as `things' to be steered, they should have the opportunity to negotiate how they are addressed/spoken to. More than telling them what they should do, they deserve options to express whether they wish to do things differently or to talk to a person who will understand.

\subsubsection{Do not assume help is needed}
Even with good intentions, assuming users need help without their explicit request (e.g., micromanaging them or proactively explaining instructions) can feel patronising for failing to support their sense of competence and dignity. 

\subsubsection{Do not treat individuals as the sum of their parts}
As applications of machine learning become more commonplace, it becomes all the more alluring to predict user preferences in terms of other users they cluster with. Whilst this may be useful to some extent, the very premise that people can be clustered into types or computationally modelled/predicted is dehumanising, as it undermines their agency as self-defining individuals. 

\subsubsection{Respect personal boundaries}
Even if a system has access to enough data to make inferences about a person, personalising responses to certain details about them can come across as invasive or ``creepy''. With the development of ubiquitous technologies, it becomes increasingly important to respect people's personal boundaries by not making them feel too ``watched'' (even if they are). 

These principles, which we aim to incorporate into a normative framework for what counts as ``good'' automated social acting, consist of what we could discern from our participants' responses. Fleshing it out will require further engagement with different demographics, in different domains, and with different modalities of social interfaces, which we aim to do in future work. 

\section{Limitations}
Among the limitations of our approach, our participant pool was limited to those aged 18-60 in English-speaking countries. We would like to repeat this in other regions and in other domains, where both cultural norms, as well as differences in system roles, could yield substantial differences in perspectives. We did not include children nor older adults in our sample, who may have very different views on the (in-)appropriate contexts for social acting. Another potential limitation stems from anchoring a large part of our investigation on smartphone notifications, which we chose for being a highly common, yet under-discussed platform behaving as a social actor. Although we asked our participants about other modalities and domains, the focus on notifications may have primed participants to focus overly on this (and, as such, the social contexts of marketing and receiving reminders to do unwanted tasks). However, notifications have not yet been analysed in this way, while the ethics of more prototypical CUIs have so far dominated debates. As we emphasised, our study was centred on \emph{negative} experiences, in order to identify anti-patterns and dark patterns, and should not be interpreted as an unbiased view of social acting in general. Finally, as with any qualitative study, there is a risk of investigator bias. We tried to reduce this risk with our phased experimental design, testing our interpretations by asking participants for clarification on comments made during earlier phases. 

\section{Conclusion}
Early CASA research has encouraged interaction designers to make systems act in increasingly socially-conforming (chatty, even friend-like) ways to elicit favourable responses from users. However, we still have a limited understanding of when the use of certain social cues in interfaces is inappropriate. This paper contributes to the literature on dark patterns/nudging by identifying and characterising an emerging `social' class of dark patterns. Drawing from a series of qualitative engagements with end-users, we also critically engage with CASA research by proposing when, and why, even seemingly innocuous automated social behaviours may bother or even offend users---constituting `social' anti-patterns. Based on end-user preferences and suggestions, we offer user-led recommendations to help interface designers prevent undesirable forms of automated social acting, and treat users in more appropriate and respectful ways. Overall, we hope our work inspires critical reflection on how automated systems treat people in interactions.


\bibliographystyle{ACM-Reference-Format}

\begin{thebibliography}{65}


\ifx \showCODEN    \undefined \def \showCODEN     #1{\unskip}     \fi
\ifx \showDOI      \undefined \def \showDOI       #1{#1}\fi
\ifx \showISBNx    \undefined \def \showISBNx     #1{\unskip}     \fi
\ifx \showISBNxiii \undefined \def \showISBNxiii  #1{\unskip}     \fi
\ifx \showISSN     \undefined \def \showISSN      #1{\unskip}     \fi
\ifx \showLCCN     \undefined \def \showLCCN      #1{\unskip}     \fi
\ifx \shownote     \undefined \def \shownote      #1{#1}          \fi
\ifx \showarticletitle \undefined \def \showarticletitle #1{#1}   \fi
\ifx \showURL      \undefined \def \showURL       {\relax}        \fi
\providecommand\bibfield[2]{#2}
\providecommand\bibinfo[2]{#2}
\providecommand\natexlab[1]{#1}
\providecommand\showeprint[2][]{arXiv:#2}

\bibitem[Abdi et~al\mbox{.}(2018)]%
        {abdi2018scoping}
\bibfield{author}{\bibinfo{person}{Jordan Abdi}, \bibinfo{person}{Ahmed Al-Hindawi}, \bibinfo{person}{Tiffany Ng}, {and} \bibinfo{person}{Marcela~P Vizcaychipi}.} \bibinfo{year}{2018}\natexlab{}.
\newblock \showarticletitle{Scoping review on the use of socially assistive robot technology in elderly care}.
\newblock \bibinfo{journal}{\emph{BMJ open}} \bibinfo{volume}{8}, \bibinfo{number}{2} (\bibinfo{year}{2018}), \bibinfo{pages}{e018815}.
\newblock


\bibitem[Amershi et~al\mbox{.}(2019)]%
        {amershi2019guidelines}
\bibfield{author}{\bibinfo{person}{Saleema Amershi}, \bibinfo{person}{Dan Weld}, \bibinfo{person}{Mihaela Vorvoreanu}, \bibinfo{person}{Adam Fourney}, \bibinfo{person}{Besmira Nushi}, \bibinfo{person}{Penny Collisson}, \bibinfo{person}{Jina Suh}, \bibinfo{person}{Shamsi Iqbal}, \bibinfo{person}{Paul~N. Bennett}, \bibinfo{person}{Kori Inkpen}, \bibinfo{person}{Jaime Teevan}, \bibinfo{person}{Ruth Kikin-Gil}, {and} \bibinfo{person}{Eric Horvitz}.} \bibinfo{year}{2019}\natexlab{}.
\newblock \showarticletitle{Guidelines for Human-AI Interaction}. In \bibinfo{booktitle}{\emph{Proceedings of the 2019 CHI Conference on Human Factors in Computing Systems}} (Glasgow, Scotland Uk) \emph{(\bibinfo{series}{CHI '19})}. \bibinfo{publisher}{Association for Computing Machinery}, \bibinfo{address}{New York, NY, USA}, \bibinfo{pages}{1–13}.
\newblock
\showISBNx{9781450359702}
\urldef\tempurl%
\url{https://doi.org/10.1145/3290605.3300233}
\showDOI{\tempurl}


\bibitem[Babushkina(2022)]%
        {babushkina2022does}
\bibfield{author}{\bibinfo{person}{Dina Babushkina}.} \bibinfo{year}{2022}\natexlab{}.
\newblock \showarticletitle{What Does It Mean for a Robot to Be Respectful?}
\newblock \bibinfo{journal}{\emph{Techn{\'e}: Research in Philosophy and Technology}} \bibinfo{volume}{26}, \bibinfo{number}{1} (\bibinfo{year}{2022}), \bibinfo{pages}{1--30}.
\newblock


\bibitem[Bender et~al\mbox{.}(2021)]%
        {bender}
\bibfield{author}{\bibinfo{person}{Emily~M. Bender}, \bibinfo{person}{Timnit Gebru}, \bibinfo{person}{Angelina McMillan-Major}, {and} \bibinfo{person}{Shmargaret Shmitchell}.} \bibinfo{year}{2021}\natexlab{}.
\newblock \showarticletitle{On the Dangers of Stochastic Parrots: Can Language Models Be Too Big?}. In \bibinfo{booktitle}{\emph{Proceedings of the 2021 ACM Conference on Fairness, Accountability, and Transparency}} (Virtual Event, Canada) \emph{(\bibinfo{series}{FAccT '21})}. \bibinfo{publisher}{Association for Computing Machinery}, \bibinfo{address}{New York, NY, USA}, \bibinfo{pages}{610–623}.
\newblock
\showISBNx{9781450383097}
\urldef\tempurl%
\url{https://doi.org/10.1145/3442188.3445922}
\showDOI{\tempurl}


\bibitem[Berens et~al\mbox{.}(2022)]%
        {7}
\bibfield{author}{\bibinfo{person}{Benjamin~Maximilian Berens}, \bibinfo{person}{Heike Dietmann}, \bibinfo{person}{Chiara Krisam}, \bibinfo{person}{Oksana Kulyk}, {and} \bibinfo{person}{Melanie Volkamer}.} \bibinfo{year}{2022}\natexlab{}.
\newblock \showarticletitle{Cookie Disclaimers: Impact of Design and Users’ Attitude}. In \bibinfo{booktitle}{\emph{Proceedings of the 17th International Conference on Availability, Reliability and Security}} (Vienna, Austria) \emph{(\bibinfo{series}{ARES '22})}. \bibinfo{publisher}{Association for Computing Machinery}, \bibinfo{address}{New York, NY, USA}, Article \bibinfo{articleno}{12}, \bibinfo{numpages}{20}~pages.
\newblock
\showISBNx{9781450396707}
\urldef\tempurl%
\url{https://doi.org/10.1145/3538969.3539008}
\showDOI{\tempurl}


\bibitem[Bermejo~Fernandez et~al\mbox{.}(2021)]%
        {9}
\bibfield{author}{\bibinfo{person}{Carlos Bermejo~Fernandez}, \bibinfo{person}{Dimitris Chatzopoulos}, \bibinfo{person}{Dimitrios Papadopoulos}, {and} \bibinfo{person}{Pan Hui}.} \bibinfo{year}{2021}\natexlab{}.
\newblock \showarticletitle{This Website Uses Nudging: MTurk Workers' Behaviour on Cookie Consent Notices}.
\newblock \bibinfo{journal}{\emph{Proc. ACM Hum.-Comput. Interact.}} \bibinfo{volume}{5}, \bibinfo{number}{CSCW2}, Article \bibinfo{articleno}{346} (\bibinfo{date}{oct} \bibinfo{year}{2021}), \bibinfo{numpages}{22}~pages.
\newblock
\urldef\tempurl%
\url{https://doi.org/10.1145/3476087}
\showDOI{\tempurl}


\bibitem[Bola{\~n}os-Castro et~al\mbox{.}(2011)]%
        {bolanos2011antipatterns}
\bibfield{author}{\bibinfo{person}{Sandro~Javier Bola{\~n}os-Castro}, \bibinfo{person}{Rub{\'e}n Gonz{\'a}lez-Crespo}, {and} \bibinfo{person}{V{\'\i}ctor~Hugo Medina~Garc{\'\i}a}.} \bibinfo{year}{2011}\natexlab{}.
\newblock \showarticletitle{Antipatterns: a compendium of bad practices in software development processes}.
\newblock \bibinfo{journal}{\emph{International Journal of Interactive Multimedia and Artificial Intelligence}} \bibinfo{volume}{1}, \bibinfo{number}{4} (\bibinfo{year}{2011}), \bibinfo{pages}{41--46}.
\newblock


\bibitem[Brabazon(2015)]%
        {brabazon2015digital}
\bibfield{author}{\bibinfo{person}{Tara Brabazon}.} \bibinfo{year}{2015}\natexlab{}.
\newblock \showarticletitle{Digital fitness: Self-monitored fitness and the commodification of movement}.
\newblock \bibinfo{journal}{\emph{Communication, Politics \& Culture}} \bibinfo{volume}{48}, \bibinfo{number}{2} (\bibinfo{year}{2015}), \bibinfo{pages}{1--23}.
\newblock


\bibitem[Braun and Clarke(2019)]%
        {braun2019reflecting}
\bibfield{author}{\bibinfo{person}{Virginia Braun} {and} \bibinfo{person}{Victoria Clarke}.} \bibinfo{year}{2019}\natexlab{}.
\newblock \showarticletitle{Reflecting on reflexive thematic analysis}.
\newblock \bibinfo{journal}{\emph{Qualitative research in sport, exercise and health}} \bibinfo{volume}{11}, \bibinfo{number}{4} (\bibinfo{year}{2019}), \bibinfo{pages}{589--597}.
\newblock


\bibitem[Breazeal et~al\mbox{.}(2016)]%
        {springer}
\bibfield{author}{\bibinfo{person}{Cynthia Breazeal}, \bibinfo{person}{Kerstin Dautenhahn}, {and} \bibinfo{person}{Takayuki Kanda}.} \bibinfo{year}{2016}\natexlab{}.
\newblock \showarticletitle{Social Robotics}.
\newblock In \bibinfo{booktitle}{\emph{Springer Handbook of Robotics}}, \bibfield{editor}{\bibinfo{person}{Bruno Siciliano} {and} \bibinfo{person}{Oussama Khatib}} (Eds.). \bibinfo{publisher}{Springer International Publishing}, \bibinfo{address}{Cham}, \bibinfo{pages}{1935--1972}.
\newblock
\showISBNx{978-3-319-32552-1}
\urldef\tempurl%
\url{https://doi.org/10.1007/978-3-319-32552-1_72}
\showDOI{\tempurl}


\bibitem[Brown et~al\mbox{.}(1998)]%
        {brown1998antipatterns}
\bibfield{author}{\bibinfo{person}{William~H. Brown}, \bibinfo{person}{Raphael~C. Malveau}, \bibinfo{person}{Hays W.~"Skip" McCormick}, {and} \bibinfo{person}{Thomas~J. Mowbray}.} \bibinfo{year}{1998}\natexlab{}.
\newblock \bibinfo{booktitle}{\emph{AntiPatterns: Refactoring Software, Architectures, and Projects in Crisis} (\bibinfo{edition}{1st} ed.)}.
\newblock \bibinfo{publisher}{John Wiley \& Sons, Inc.}, \bibinfo{address}{USA}.
\newblock
\showISBNx{0471197130}


\bibitem[Caraban et~al\mbox{.}(2019)]%
        {caraban201923}
\bibfield{author}{\bibinfo{person}{Ana Caraban}, \bibinfo{person}{Evangelos Karapanos}, \bibinfo{person}{Daniel Gon\c{c}alves}, {and} \bibinfo{person}{Pedro Campos}.} \bibinfo{year}{2019}\natexlab{}.
\newblock \showarticletitle{23 Ways to Nudge: A Review of Technology-Mediated Nudging in Human-Computer Interaction}. In \bibinfo{booktitle}{\emph{Proceedings of the 2019 CHI Conference on Human Factors in Computing Systems}} (Glasgow, Scotland Uk) \emph{(\bibinfo{series}{CHI '19})}. \bibinfo{publisher}{Association for Computing Machinery}, \bibinfo{address}{New York, NY, USA}, \bibinfo{pages}{1–15}.
\newblock
\showISBNx{9781450359702}
\urldef\tempurl%
\url{https://doi.org/10.1145/3290605.3300733}
\showDOI{\tempurl}


\bibitem[Chun~Tie et~al\mbox{.}(2019)]%
        {chun2019grounded}
\bibfield{author}{\bibinfo{person}{Ylona Chun~Tie}, \bibinfo{person}{Melanie Birks}, {and} \bibinfo{person}{Karen Francis}.} \bibinfo{year}{2019}\natexlab{}.
\newblock \showarticletitle{Grounded theory research: A design framework for novice researchers}.
\newblock \bibinfo{journal}{\emph{SAGE open medicine}}  \bibinfo{volume}{7} (\bibinfo{year}{2019}), \bibinfo{pages}{2050312118822927}.
\newblock


\bibitem[Ciechanowski et~al\mbox{.}(2019)]%
        {6}
\bibfield{author}{\bibinfo{person}{Leon Ciechanowski}, \bibinfo{person}{Aleksandra Przegalinska}, \bibinfo{person}{Mikolaj Magnuski}, {and} \bibinfo{person}{Peter Gloor}.} \bibinfo{year}{2019}\natexlab{}.
\newblock \showarticletitle{In the shades of the uncanny valley: An experimental study of human--chatbot interaction}.
\newblock \bibinfo{journal}{\emph{Future Generation Computer Systems}}  \bibinfo{volume}{92} (\bibinfo{year}{2019}), \bibinfo{pages}{539--548}.
\newblock


\bibitem[Clarke et~al\mbox{.}(2015)]%
        {clarke2015thematic}
\bibfield{author}{\bibinfo{person}{Victoria Clarke}, \bibinfo{person}{Virginia Braun}, {and} \bibinfo{person}{Nikki Hayfield}.} \bibinfo{year}{2015}\natexlab{}.
\newblock \showarticletitle{Thematic analysis}.
\newblock \bibinfo{journal}{\emph{Qualitative psychology: A practical guide to research methods}} \bibinfo{volume}{222}, \bibinfo{number}{2015} (\bibinfo{year}{2015}), \bibinfo{pages}{248}.
\newblock


\bibitem[Conti and Sobiesk(2010)]%
        {conti2010malicious}
\bibfield{author}{\bibinfo{person}{Gregory Conti} {and} \bibinfo{person}{Edward Sobiesk}.} \bibinfo{year}{2010}\natexlab{}.
\newblock \showarticletitle{Malicious Interface Design: Exploiting the User}. In \bibinfo{booktitle}{\emph{Proceedings of the 19th International Conference on World Wide Web}} (Raleigh, North Carolina, USA) \emph{(\bibinfo{series}{WWW '10})}. \bibinfo{publisher}{Association for Computing Machinery}, \bibinfo{address}{New York, NY, USA}, \bibinfo{pages}{271–280}.
\newblock
\showISBNx{9781605587998}
\urldef\tempurl%
\url{https://doi.org/10.1145/1772690.1772719}
\showDOI{\tempurl}


\bibitem[Cranor(2022)]%
        {281}
\bibfield{author}{\bibinfo{person}{Lorrie~Faith Cranor}.} \bibinfo{year}{2022}\natexlab{}.
\newblock \showarticletitle{Cookie Monster}.
\newblock \bibinfo{journal}{\emph{Commun. ACM}} \bibinfo{volume}{65}, \bibinfo{number}{7} (\bibinfo{date}{jun} \bibinfo{year}{2022}), \bibinfo{pages}{30–32}.
\newblock
\showISSN{0001-0782}
\urldef\tempurl%
\url{https://doi.org/10.1145/3538639}
\showDOI{\tempurl}


\bibitem[Cross et~al\mbox{.}(2016)]%
        {cross2016shaping}
\bibfield{author}{\bibinfo{person}{Emily~S Cross}, \bibinfo{person}{Richard Ramsey}, \bibinfo{person}{Roman Liepelt}, \bibinfo{person}{Wolfgang Prinz}, {and} \bibinfo{person}{Antonia F de~C Hamilton}.} \bibinfo{year}{2016}\natexlab{}.
\newblock \showarticletitle{The shaping of social perception by stimulus and knowledge cues to human animacy}.
\newblock \bibinfo{journal}{\emph{Philosophical Transactions of the Royal Society B: Biological Sciences}} \bibinfo{volume}{371}, \bibinfo{number}{1686} (\bibinfo{year}{2016}), \bibinfo{pages}{20150075}.
\newblock


\bibitem[Dodsworth(2021)]%
        {dodsworth2021state}
\bibfield{author}{\bibinfo{person}{Laura Dodsworth}.} \bibinfo{year}{2021}\natexlab{}.
\newblock \bibinfo{booktitle}{\emph{A state of fear: How the UK government weaponised fear during the Covid-19 pandemic}}.
\newblock \bibinfo{publisher}{Pinter \& Martin}, \bibinfo{address}{London, UK}.
\newblock


\bibitem[Dolan et~al\mbox{.}(2010)]%
        {dolan2010mindspace}
\bibfield{author}{\bibinfo{person}{Paul Dolan}, \bibinfo{person}{Michael Hallsworth}, \bibinfo{person}{David Halpern}, \bibinfo{person}{Dominic King}, {and} \bibinfo{person}{Ivo Vlaev}.} \bibinfo{year}{2010}\natexlab{}.
\newblock \bibinfo{booktitle}{\emph{MINDSPACE: influencing behaviour for public policy}}.
\newblock \bibinfo{type}{{T}echnical {R}eport}. \bibinfo{institution}{Institute of Government}.
\newblock


\bibitem[Donath(2020)]%
        {donath2020ethical}
\bibfield{author}{\bibinfo{person}{Judith Donath}.} \bibinfo{year}{2020}\natexlab{}.
\newblock \showarticletitle{{5253Ethical Issues in Our Relationship with Artificial Entities}}.
\newblock In \bibinfo{booktitle}{\emph{{The Oxford Handbook of Ethics of AI}}}. \bibinfo{publisher}{Oxford University Press}, \bibinfo{address}{Oxford, UK}.
\newblock
\showISBNx{9780190067397}
\urldef\tempurl%
\url{https://doi.org/10.1093/oxfordhb/9780190067397.013.3}
\showDOI{\tempurl}
\showeprint{https://academic.oup.com/book/0/chapter/290656277/chapter-ag-pdf/44521956/book\_34287\_section\_290656277.ag.pdf}


\bibitem[F{\o}lstad and Brandtzaeg(2020)]%
        {folstad2020users}
\bibfield{author}{\bibinfo{person}{Asbj{\o}rn F{\o}lstad} {and} \bibinfo{person}{Petter~Bae Brandtzaeg}.} \bibinfo{year}{2020}\natexlab{}.
\newblock \showarticletitle{Users' experiences with chatbots: findings from a questionnaire study}.
\newblock \bibinfo{journal}{\emph{Quality and User Experience}} \bibinfo{volume}{5}, \bibinfo{number}{1} (\bibinfo{year}{2020}), \bibinfo{pages}{1--14}.
\newblock


\bibitem[Gambino et~al\mbox{.}(2020)]%
        {gambino2020building}
\bibfield{author}{\bibinfo{person}{Andrew Gambino}, \bibinfo{person}{Jesse Fox}, {and} \bibinfo{person}{Rabindra~A Ratan}.} \bibinfo{year}{2020}\natexlab{}.
\newblock \showarticletitle{Building a stronger CASA: Extending the computers are social actors paradigm}.
\newblock \bibinfo{journal}{\emph{Human-Machine Communication}} \bibinfo{volume}{1}, \bibinfo{number}{1} (\bibinfo{year}{2020}), \bibinfo{pages}{5}.
\newblock


\bibitem[Gray et~al\mbox{.}(2018)]%
        {gray2018dark}
\bibfield{author}{\bibinfo{person}{Colin~M. Gray}, \bibinfo{person}{Yubo Kou}, \bibinfo{person}{Bryan Battles}, \bibinfo{person}{Joseph Hoggatt}, {and} \bibinfo{person}{Austin~L. Toombs}.} \bibinfo{year}{2018}\natexlab{}.
\newblock \showarticletitle{The Dark (Patterns) Side of UX Design}. In \bibinfo{booktitle}{\emph{Proceedings of the 2018 CHI Conference on Human Factors in Computing Systems}} (Montreal QC, Canada) \emph{(\bibinfo{series}{CHI '18})}. \bibinfo{publisher}{Association for Computing Machinery}, \bibinfo{address}{New York, NY, USA}, \bibinfo{pages}{1–14}.
\newblock
\showISBNx{9781450356206}
\urldef\tempurl%
\url{https://doi.org/10.1145/3173574.3174108}
\showDOI{\tempurl}


\bibitem[Greenberg et~al\mbox{.}(2014)]%
        {proxemic}
\bibfield{author}{\bibinfo{person}{Saul Greenberg}, \bibinfo{person}{Sebastian Boring}, \bibinfo{person}{Jo Vermeulen}, {and} \bibinfo{person}{Jakub Dostal}.} \bibinfo{year}{2014}\natexlab{}.
\newblock \showarticletitle{Dark Patterns in Proxemic Interactions: A Critical Perspective}. In \bibinfo{booktitle}{\emph{Proceedings of the 2014 Conference on Designing Interactive Systems}} (Vancouver, BC, Canada) \emph{(\bibinfo{series}{DIS '14})}. \bibinfo{publisher}{Association for Computing Machinery}, \bibinfo{address}{New York, NY, USA}, \bibinfo{pages}{523–532}.
\newblock
\showISBNx{9781450329026}
\urldef\tempurl%
\url{https://doi.org/10.1145/2598510.2598541}
\showDOI{\tempurl}


\bibitem[Grimes et~al\mbox{.}(2021)]%
        {grimes2021mental}
\bibfield{author}{\bibinfo{person}{G~Mark Grimes}, \bibinfo{person}{Ryan~M Schuetzler}, {and} \bibinfo{person}{Justin~Scott Giboney}.} \bibinfo{year}{2021}\natexlab{}.
\newblock \showarticletitle{Mental models and expectation violations in conversational AI interactions}.
\newblock \bibinfo{journal}{\emph{Decision Support Systems}}  \bibinfo{volume}{144} (\bibinfo{year}{2021}), \bibinfo{pages}{113515}.
\newblock


\bibitem[Habib et~al\mbox{.}(2022)]%
        {soc}
\bibfield{author}{\bibinfo{person}{Hana Habib}, \bibinfo{person}{Sarah Pearman}, \bibinfo{person}{Ellie Young}, \bibinfo{person}{Ishika Saxena}, \bibinfo{person}{Robert Zhang}, {and} \bibinfo{person}{Lorrie~FaIth Cranor}.} \bibinfo{year}{2022}\natexlab{}.
\newblock \showarticletitle{Identifying User Needs for Advertising Controls on Facebook}.
\newblock \bibinfo{journal}{\emph{Proc. ACM Hum.-Comput. Interact.}} \bibinfo{volume}{6}, \bibinfo{number}{CSCW1}, Article \bibinfo{articleno}{59} (\bibinfo{date}{apr} \bibinfo{year}{2022}), \bibinfo{numpages}{42}~pages.
\newblock
\urldef\tempurl%
\url{https://doi.org/10.1145/3512906}
\showDOI{\tempurl}


\bibitem[Hadi(2019)]%
        {hadi2019humanizing}
\bibfield{author}{\bibinfo{person}{Rhonda Hadi}.} \bibinfo{year}{2019}\natexlab{}.
\newblock \showarticletitle{When Humanizing Customer Service Chatbots Might Backfire}.
\newblock \bibinfo{journal}{\emph{NIM Marketing Intelligence Review}} \bibinfo{volume}{11}, \bibinfo{number}{2} (\bibinfo{year}{2019}), \bibinfo{pages}{30--35}.
\newblock
\urldef\tempurl%
\url{https://doi.org/10.2478/nimmir-2019-0013}
\showDOI{\tempurl}


\bibitem[Kocielnik et~al\mbox{.}(2021)]%
        {kocielnik2021can}
\bibfield{author}{\bibinfo{person}{Rafal Kocielnik}, \bibinfo{person}{Raina Langevin}, \bibinfo{person}{James~S. George}, \bibinfo{person}{Shota Akenaga}, \bibinfo{person}{Amelia Wang}, \bibinfo{person}{Darwin~P. Jones}, \bibinfo{person}{Alexander Argyle}, \bibinfo{person}{Callan Fockele}, \bibinfo{person}{Layla Anderson}, \bibinfo{person}{Dennis~T. Hsieh}, \bibinfo{person}{Kabir Yadav}, \bibinfo{person}{Herbert Duber}, \bibinfo{person}{Gary Hsieh}, {and} \bibinfo{person}{Andrea~L. Hartzler}.} \bibinfo{year}{2021}\natexlab{}.
\newblock \showarticletitle{Can I Talk to You about Your Social Needs? Understanding Preference for Conversational User Interface in Health}. In \bibinfo{booktitle}{\emph{Proceedings of the 3rd Conference on Conversational User Interfaces}} (Bilbao (online), Spain) \emph{(\bibinfo{series}{CUI '21})}. \bibinfo{publisher}{Association for Computing Machinery}, \bibinfo{address}{New York, NY, USA}, Article \bibinfo{articleno}{4}, \bibinfo{numpages}{10}~pages.
\newblock
\showISBNx{9781450389983}
\urldef\tempurl%
\url{https://doi.org/10.1145/3469595.3469599}
\showDOI{\tempurl}


\bibitem[Lacey and Caudwell(2019)]%
        {cute}
\bibfield{author}{\bibinfo{person}{Cherie Lacey} {and} \bibinfo{person}{Catherine Caudwell}.} \bibinfo{year}{2019}\natexlab{}.
\newblock \showarticletitle{Cuteness as a ‘Dark Pattern’ in Home Robots}. In \bibinfo{booktitle}{\emph{14th ACM/IEEE International Conference on Human-Robot Interaction (HRI)}}. \bibinfo{publisher}{IEEE}, \bibinfo{address}{Daegu, Korea (South)}, \bibinfo{pages}{374--381}.
\newblock
\urldef\tempurl%
\url{https://doi.org/10.1109/HRI.2019.8673274}
\showDOI{\tempurl}


\bibitem[Laitinen et~al\mbox{.}(2016)]%
        {laitinen2016social}
\bibfield{author}{\bibinfo{person}{Arto Laitinen}, \bibinfo{person}{Marketta Niemel{\"a}}, {and} \bibinfo{person}{Jari Pirhonen}.} \bibinfo{year}{2016}\natexlab{}.
\newblock \showarticletitle{Social robotics, elderly care, and human dignity: a recognition-theoretical approach}.
\newblock In \bibinfo{booktitle}{\emph{What social robots can and should do}}. \bibinfo{publisher}{IOS Press}, \bibinfo{address}{Amsterdam, The Netherlands}, \bibinfo{pages}{155--163}.
\newblock


\bibitem[Lee(2010)]%
        {lee2010more}
\bibfield{author}{\bibinfo{person}{Eun-Ju Lee}.} \bibinfo{year}{2010}\natexlab{}.
\newblock \showarticletitle{The more humanlike, the better? How speech type and users’ cognitive style affect social responses to computers}.
\newblock \bibinfo{journal}{\emph{Computers in Human Behavior}} \bibinfo{volume}{26}, \bibinfo{number}{4} (\bibinfo{year}{2010}), \bibinfo{pages}{665--672}.
\newblock


\bibitem[Leonard(2008)]%
        {nudge}
\bibfield{author}{\bibinfo{person}{Thomas~C Leonard}.} \bibinfo{year}{2008}\natexlab{}.
\newblock \bibinfo{booktitle}{\emph{Richard H. Thaler, Cass R. Sunstein, Nudge: Improving decisions about health, wealth, and happiness}}.
\newblock \bibinfo{publisher}{Yale University Press}, \bibinfo{address}{New Haven, CT}.
\newblock


\bibitem[Li et~al\mbox{.}(2023)]%
        {li2022assessing}
\bibfield{author}{\bibinfo{person}{Tianyi Li}, \bibinfo{person}{Mihaela Vorvoreanu}, \bibinfo{person}{Derek Debellis}, {and} \bibinfo{person}{Saleema Amershi}.} \bibinfo{year}{2023}\natexlab{}.
\newblock \showarticletitle{Assessing Human-AI Interaction Early through Factorial Surveys: A Study on the Guidelines for Human-AI Interaction}.
\newblock \bibinfo{journal}{\emph{ACM Trans. Comput.-Hum. Interact.}} \bibinfo{volume}{30}, \bibinfo{number}{5}, Article \bibinfo{articleno}{69} (\bibinfo{date}{sep} \bibinfo{year}{2023}), \bibinfo{numpages}{45}~pages.
\newblock
\showISSN{1073-0516}
\urldef\tempurl%
\url{https://doi.org/10.1145/3511605}
\showDOI{\tempurl}


\bibitem[Long(2001)]%
        {long2003development}
\bibfield{author}{\bibinfo{person}{Norman Long}.} \bibinfo{year}{2001}\natexlab{}.
\newblock \bibinfo{booktitle}{\emph{Development sociology: actor perspectives}}.
\newblock \bibinfo{publisher}{Routledge}, \bibinfo{address}{New York, NY}.
\newblock


\bibitem[Lucas et~al\mbox{.}(2017)]%
        {lucas2017role}
\bibfield{author}{\bibinfo{person}{Gale~M. Lucas}, \bibinfo{person}{Jill Boberg}, \bibinfo{person}{David Traum}, \bibinfo{person}{Ron Artstein}, \bibinfo{person}{Jon Gratch}, \bibinfo{person}{Alesia Gainer}, \bibinfo{person}{Emmanuel Johnson}, \bibinfo{person}{Anton Leuski}, {and} \bibinfo{person}{Mikio Nakano}.} \bibinfo{year}{2017}\natexlab{}.
\newblock \showarticletitle{The Role of Social Dialogue and Errors in Robots}. In \bibinfo{booktitle}{\emph{Proceedings of the 5th International Conference on Human Agent Interaction}} (Bielefeld, Germany) \emph{(\bibinfo{series}{HAI '17})}. \bibinfo{publisher}{Association for Computing Machinery}, \bibinfo{address}{New York, NY, USA}, \bibinfo{pages}{431–433}.
\newblock
\showISBNx{9781450351133}
\urldef\tempurl%
\url{https://doi.org/10.1145/3125739.3132617}
\showDOI{\tempurl}


\bibitem[Luger and Sellen(2016)]%
        {33}
\bibfield{author}{\bibinfo{person}{Ewa Luger} {and} \bibinfo{person}{Abigail Sellen}.} \bibinfo{year}{2016}\natexlab{}.
\newblock \showarticletitle{"Like Having a Really Bad PA": The Gulf between User Expectation and Experience of Conversational Agents}. In \bibinfo{booktitle}{\emph{Proceedings of the 2016 CHI Conference on Human Factors in Computing Systems}} (San Jose, California, USA) \emph{(\bibinfo{series}{CHI '16})}. \bibinfo{publisher}{Association for Computing Machinery}, \bibinfo{address}{New York, NY, USA}, \bibinfo{pages}{5286–5297}.
\newblock
\showISBNx{9781450333627}
\urldef\tempurl%
\url{https://doi.org/10.1145/2858036.2858288}
\showDOI{\tempurl}


\bibitem[Lukoff et~al\mbox{.}(2021a)]%
        {lukoff2021can}
\bibfield{author}{\bibinfo{person}{Kai Lukoff}, \bibinfo{person}{Alexis Hiniker}, \bibinfo{person}{Colin~M. Gray}, \bibinfo{person}{Arunesh Mathur}, {and} \bibinfo{person}{Shruthi~Sai Chivukula}.} \bibinfo{year}{2021}\natexlab{a}.
\newblock \showarticletitle{What Can CHI Do About Dark Patterns?}. In \bibinfo{booktitle}{\emph{Extended Abstracts of the 2021 CHI Conference on Human Factors in Computing Systems}} (Yokohama, Japan) \emph{(\bibinfo{series}{CHI EA '21})}. \bibinfo{publisher}{Association for Computing Machinery}, \bibinfo{address}{New York, NY, USA}, Article \bibinfo{articleno}{102}, \bibinfo{numpages}{6}~pages.
\newblock
\showISBNx{9781450380959}
\urldef\tempurl%
\url{https://doi.org/10.1145/3411763.3441360}
\showDOI{\tempurl}


\bibitem[Lukoff et~al\mbox{.}(2021b)]%
        {lukoff2021design}
\bibfield{author}{\bibinfo{person}{Kai Lukoff}, \bibinfo{person}{Ulrik Lyngs}, \bibinfo{person}{Himanshu Zade}, \bibinfo{person}{J.~Vera Liao}, \bibinfo{person}{James Choi}, \bibinfo{person}{Kaiyue Fan}, \bibinfo{person}{Sean~A. Munson}, {and} \bibinfo{person}{Alexis Hiniker}.} \bibinfo{year}{2021}\natexlab{b}.
\newblock \showarticletitle{How the Design of YouTube Influences User Sense of Agency}. In \bibinfo{booktitle}{\emph{Proceedings of the 2021 CHI Conference on Human Factors in Computing Systems}} (Yokohama, Japan) \emph{(\bibinfo{series}{CHI '21})}. \bibinfo{publisher}{Association for Computing Machinery}, \bibinfo{address}{New York, NY, USA}, Article \bibinfo{articleno}{368}, \bibinfo{numpages}{17}~pages.
\newblock
\showISBNx{9781450380966}
\urldef\tempurl%
\url{https://doi.org/10.1145/3411764.3445467}
\showDOI{\tempurl}


\bibitem[MacDonald(2019)]%
        {macdonald2019anti}
\bibfield{author}{\bibinfo{person}{Diana MacDonald}.} \bibinfo{year}{2019}\natexlab{}.
\newblock \bibinfo{booktitle}{\emph{Anti-patterns and dark patterns}}.
\newblock \bibinfo{publisher}{Apress}, \bibinfo{address}{Berkeley, CA}, \bibinfo{pages}{193--221}.
\newblock
\showISBNx{978-1-4842-4938-3}
\urldef\tempurl%
\url{https://doi.org/10.1007/978-1-4842-4938-3_5}
\showDOI{\tempurl}


\bibitem[Mathur et~al\mbox{.}(2019)]%
        {mathur2019dark}
\bibfield{author}{\bibinfo{person}{Arunesh Mathur}, \bibinfo{person}{Gunes Acar}, \bibinfo{person}{Michael~J. Friedman}, \bibinfo{person}{Eli Lucherini}, \bibinfo{person}{Jonathan Mayer}, \bibinfo{person}{Marshini Chetty}, {and} \bibinfo{person}{Arvind Narayanan}.} \bibinfo{year}{2019}\natexlab{}.
\newblock \showarticletitle{Dark Patterns at Scale: Findings from a Crawl of 11K Shopping Websites}.
\newblock \bibinfo{journal}{\emph{Proc. ACM Hum.-Comput. Interact.}} \bibinfo{volume}{3}, \bibinfo{number}{CSCW}, Article \bibinfo{articleno}{81} (\bibinfo{date}{nov} \bibinfo{year}{2019}), \bibinfo{numpages}{32}~pages.
\newblock
\urldef\tempurl%
\url{https://doi.org/10.1145/3359183}
\showDOI{\tempurl}


\bibitem[Mathur et~al\mbox{.}(2021)]%
        {mathur21}
\bibfield{author}{\bibinfo{person}{Arunesh Mathur}, \bibinfo{person}{Mihir Kshirsagar}, {and} \bibinfo{person}{Jonathan Mayer}.} \bibinfo{year}{2021}\natexlab{}.
\newblock \showarticletitle{What Makes a Dark Pattern... Dark? Design Attributes, Normative Considerations, and Measurement Methods}. In \bibinfo{booktitle}{\emph{Proceedings of the 2021 CHI Conference on Human Factors in Computing Systems}} (Yokohama, Japan) \emph{(\bibinfo{series}{CHI '21})}. \bibinfo{publisher}{Association for Computing Machinery}, \bibinfo{address}{New York, NY, USA}, Article \bibinfo{articleno}{360}, \bibinfo{numpages}{18}~pages.
\newblock
\showISBNx{9781450380966}
\urldef\tempurl%
\url{https://doi.org/10.1145/3411764.3445610}
\showDOI{\tempurl}


\bibitem[Meng and Dai(2021)]%
        {meng2021emotional}
\bibfield{author}{\bibinfo{person}{Jingbo Meng} {and} \bibinfo{person}{Yue~(Nancy) Dai}.} \bibinfo{year}{2021}\natexlab{}.
\newblock \showarticletitle{{Emotional Support from AI Chatbots: Should a Supportive Partner Self-Disclose or Not?}}
\newblock \bibinfo{journal}{\emph{Journal of Computer-Mediated Communication}} \bibinfo{volume}{26}, \bibinfo{number}{4} (\bibinfo{date}{05} \bibinfo{year}{2021}), \bibinfo{pages}{207--222}.
\newblock
\showISSN{1083-6101}
\urldef\tempurl%
\url{https://doi.org/10.1093/jcmc/zmab005}
\showDOI{\tempurl}
\showeprint{https://academic.oup.com/jcmc/article-pdf/26/4/207/40342390/zmab005.pdf}


\bibitem[Moser et~al\mbox{.}(2019)]%
        {com}
\bibfield{author}{\bibinfo{person}{Carol Moser}, \bibinfo{person}{Sarita~Y. Schoenebeck}, {and} \bibinfo{person}{Paul Resnick}.} \bibinfo{year}{2019}\natexlab{}.
\newblock \showarticletitle{Impulse Buying: Design Practices and Consumer Needs}. In \bibinfo{booktitle}{\emph{Proceedings of the 2019 CHI Conference on Human Factors in Computing Systems}} (Glasgow, Scotland Uk) \emph{(\bibinfo{series}{CHI '19})}. \bibinfo{publisher}{Association for Computing Machinery}, \bibinfo{address}{New York, NY, USA}, \bibinfo{pages}{1–15}.
\newblock
\showISBNx{9781450359702}
\urldef\tempurl%
\url{https://doi.org/10.1145/3290605.3300472}
\showDOI{\tempurl}


\bibitem[Narayanan et~al\mbox{.}(2020)]%
        {narayanan2020dark}
\bibfield{author}{\bibinfo{person}{Arvind Narayanan}, \bibinfo{person}{Arunesh Mathur}, \bibinfo{person}{Marshini Chetty}, {and} \bibinfo{person}{Mihir Kshirsagar}.} \bibinfo{year}{2020}\natexlab{}.
\newblock \showarticletitle{Dark Patterns: Past, Present, and Future: The evolution of tricky user interfaces}.
\newblock \bibinfo{journal}{\emph{Queue}} \bibinfo{volume}{18}, \bibinfo{number}{2} (\bibinfo{year}{2020}), \bibinfo{pages}{67--92}.
\newblock


\bibitem[Nass and Moon(2000)]%
        {nass2000machines}
\bibfield{author}{\bibinfo{person}{Clifford Nass} {and} \bibinfo{person}{Youngme Moon}.} \bibinfo{year}{2000}\natexlab{}.
\newblock \showarticletitle{Machines and mindlessness: Social responses to computers}.
\newblock \bibinfo{journal}{\emph{Journal of social issues}} \bibinfo{volume}{56}, \bibinfo{number}{1} (\bibinfo{year}{2000}), \bibinfo{pages}{81--103}.
\newblock


\bibitem[Nass et~al\mbox{.}(1994)]%
        {nass1994computers}
\bibfield{author}{\bibinfo{person}{Clifford Nass}, \bibinfo{person}{Jonathan Steuer}, {and} \bibinfo{person}{Ellen~R. Tauber}.} \bibinfo{year}{1994}\natexlab{}.
\newblock \showarticletitle{Computers Are Social Actors}. In \bibinfo{booktitle}{\emph{Proceedings of the SIGCHI Conference on Human Factors in Computing Systems}} (Boston, Massachusetts, USA) \emph{(\bibinfo{series}{CHI '94})}. \bibinfo{publisher}{Association for Computing Machinery}, \bibinfo{address}{New York, NY, USA}, \bibinfo{pages}{72–78}.
\newblock
\showISBNx{0897916506}
\urldef\tempurl%
\url{https://doi.org/10.1145/191666.191703}
\showDOI{\tempurl}


\bibitem[O'Reilly et~al\mbox{.}(2021)]%
        {o2021mixing}
\bibfield{author}{\bibinfo{person}{Michelle O'Reilly}, \bibinfo{person}{Nikki Kiyimba}, {and} \bibinfo{person}{Alison Drewett}.} \bibinfo{year}{2021}\natexlab{}.
\newblock \showarticletitle{Mixing qualitative methods versus methodologies: A critical reflection on communication and power in inpatient care}.
\newblock \bibinfo{journal}{\emph{Counselling and Psychotherapy Research}} \bibinfo{volume}{21}, \bibinfo{number}{1} (\bibinfo{year}{2021}), \bibinfo{pages}{66--76}.
\newblock


\bibitem[Owens et~al\mbox{.}(2022)]%
        {owens2022exploring}
\bibfield{author}{\bibinfo{person}{Kentrell Owens}, \bibinfo{person}{Johanna Gunawan}, \bibinfo{person}{David Choffnes}, \bibinfo{person}{Pardis Emami-Naeini}, \bibinfo{person}{Tadayoshi Kohno}, {and} \bibinfo{person}{Franziska Roesner}.} \bibinfo{year}{2022}\natexlab{}.
\newblock \showarticletitle{Exploring Deceptive Design Patterns in Voice Interfaces}. In \bibinfo{booktitle}{\emph{Proceedings of the 2022 European Symposium on Usable Security}} (Karlsruhe, Germany) \emph{(\bibinfo{series}{EuroUSEC '22})}. \bibinfo{publisher}{Association for Computing Machinery}, \bibinfo{address}{New York, NY, USA}, \bibinfo{pages}{64–78}.
\newblock
\showISBNx{9781450397001}
\urldef\tempurl%
\url{https://doi.org/10.1145/3549015.3554213}
\showDOI{\tempurl}


\bibitem[O’Reilly and Kiyimba(2015)]%
        {o2015advanced}
\bibfield{author}{\bibinfo{person}{Michelle O’Reilly} {and} \bibinfo{person}{Nikki Kiyimba}.} \bibinfo{year}{2015}\natexlab{}.
\newblock \bibinfo{booktitle}{\emph{Advanced qualitative research: A guide to using theory}}.
\newblock \bibinfo{publisher}{Sage}, \bibinfo{address}{New York, NY}.
\newblock


\bibitem[Pan et~al\mbox{.}(2023)]%
        {pan2023automatically}
\bibfield{author}{\bibinfo{person}{Liangming Pan}, \bibinfo{person}{Michael Saxon}, \bibinfo{person}{Wenda Xu}, \bibinfo{person}{Deepak Nathani}, \bibinfo{person}{Xinyi Wang}, {and} \bibinfo{person}{William~Yang Wang}.} \bibinfo{year}{2023}\natexlab{}.
\newblock \bibinfo{title}{Automatically Correcting Large Language Models: Surveying the landscape of diverse self-correction strategies}.
\newblock
\newblock
\showeprint[arxiv]{2308.03188}~[cs.CL]


\bibitem[Reeves and Nass(1996)]%
        {reeves1996media}
\bibfield{author}{\bibinfo{person}{Byron Reeves} {and} \bibinfo{person}{Clifford Nass}.} \bibinfo{year}{1996}\natexlab{}.
\newblock \bibinfo{booktitle}{\emph{The media equation: How people treat computers, television, and new media like real people}}.
\newblock \bibinfo{publisher}{Cambridge university press}, \bibinfo{address}{Cambridge, UK}.
\newblock


\bibitem[Seeger et~al\mbox{.}(2017)]%
        {seeger2017we}
\bibfield{author}{\bibinfo{person}{Anna-Maria Seeger}, \bibinfo{person}{Jella Pfeiffer}, {and} \bibinfo{person}{Armin Heinzl}.} \bibinfo{year}{2017}\natexlab{}.
\newblock \showarticletitle{When do we need a human? Anthropomorphic design and trustworthiness of conversational agents}. In \bibinfo{booktitle}{\emph{SIGHCI 2017 Proceedings}}. \bibinfo{publisher}{Association for Information Systems}, \bibinfo{address}{Seoul, Korea}, \bibinfo{numpages}{15}~pages.
\newblock


\bibitem[Shamsudhin and Jotterand(2021)]%
        {shamsudhin2022social}
\bibfield{author}{\bibinfo{person}{Naveen Shamsudhin} {and} \bibinfo{person}{Fabrice Jotterand}.} \bibinfo{year}{2021}\natexlab{}.
\newblock \showarticletitle{Social Robots and Dark Patterns: Where Does Persuasion End and Deception Begin?}
\newblock In \bibinfo{booktitle}{\emph{Artificial Intelligence in Brain and Mental Health: Philosophical, Ethical {\&} Policy Issues}}, \bibfield{editor}{\bibinfo{person}{Fabrice Jotterand} {and} \bibinfo{person}{Marcello Ienca}} (Eds.). \bibinfo{publisher}{Springer International Publishing}, \bibinfo{address}{Cham}, \bibinfo{pages}{89--110}.
\newblock
\showISBNx{978-3-030-74188-4}
\urldef\tempurl%
\url{https://doi.org/10.1007/978-3-030-74188-4_7}
\showDOI{\tempurl}


\bibitem[Shelby et~al\mbox{.}(2023)]%
        {Renee}
\bibfield{author}{\bibinfo{person}{Renee Shelby}, \bibinfo{person}{Shalaleh Rismani}, \bibinfo{person}{Kathryn Henne}, \bibinfo{person}{AJung Moon}, \bibinfo{person}{Negar Rostamzadeh}, \bibinfo{person}{Paul Nicholas}, \bibinfo{person}{N'Mah Yilla-Akbari}, \bibinfo{person}{Jess Gallegos}, \bibinfo{person}{Andrew Smart}, \bibinfo{person}{Emilio Garcia}, {and} \bibinfo{person}{Gurleen Virk}.} \bibinfo{year}{2023}\natexlab{}.
\newblock \showarticletitle{Sociotechnical Harms of Algorithmic Systems: Scoping a Taxonomy for Harm Reduction}. In \bibinfo{booktitle}{\emph{Proceedings of the 2023 AAAI/ACM Conference on AI, Ethics, and Society}} (Montr\'{e}al, QC, Canada) \emph{(\bibinfo{series}{AIES '23})}. \bibinfo{publisher}{Association for Computing Machinery}, \bibinfo{address}{New York, NY, USA}, \bibinfo{pages}{723–741}.
\newblock
\showISBNx{9798400702310}
\urldef\tempurl%
\url{https://doi.org/10.1145/3600211.3604673}
\showDOI{\tempurl}


\bibitem[Sin et~al\mbox{.}(2022)]%
        {comm}
\bibfield{author}{\bibinfo{person}{Ray Sin}, \bibinfo{person}{Ted Harris}, \bibinfo{person}{Simon Nilsson}, {and} \bibinfo{person}{Talia Beck}.} \bibinfo{year}{2022}\natexlab{}.
\newblock \showarticletitle{Dark patterns in online shopping: do they work and can nudges help mitigate impulse buying?}
\newblock \bibinfo{journal}{\emph{Behavioural Public Policy}} (\bibinfo{year}{2022}), \bibinfo{pages}{1–27}.
\newblock
\urldef\tempurl%
\url{https://doi.org/10.1017/bpp.2022.11}
\showDOI{\tempurl}


\bibitem[Suchman(2007)]%
        {suchman2007human}
\bibfield{author}{\bibinfo{person}{Lucy Suchman}.} \bibinfo{year}{2007}\natexlab{}.
\newblock \bibinfo{booktitle}{\emph{Human-machine reconfigurations: Plans and situated actions}}.
\newblock \bibinfo{publisher}{Cambridge University Press}, \bibinfo{address}{Cambridge, UK}.
\newblock


\bibitem[Svenningsson and Faraon(2020)]%
        {svenningsson2019artificial}
\bibfield{author}{\bibinfo{person}{Nina Svenningsson} {and} \bibinfo{person}{Montathar Faraon}.} \bibinfo{year}{2020}\natexlab{}.
\newblock \showarticletitle{Artificial Intelligence in Conversational Agents: A Study of Factors Related to Perceived Humanness in Chatbots}. In \bibinfo{booktitle}{\emph{Proceedings of the 2019 2nd Artificial Intelligence and Cloud Computing Conference}} (Kobe, Japan) \emph{(\bibinfo{series}{AICCC 2019})}. \bibinfo{publisher}{Association for Computing Machinery}, \bibinfo{address}{New York, NY, USA}, \bibinfo{pages}{151–161}.
\newblock
\showISBNx{9781450372633}
\urldef\tempurl%
\url{https://doi.org/10.1145/3375959.3375973}
\showDOI{\tempurl}


\bibitem[Svikhnushina et~al\mbox{.}(2021)]%
        {svikhnushina2020social}
\bibfield{author}{\bibinfo{person}{Ekaterina Svikhnushina}, \bibinfo{person}{Alexandru Placinta}, {and} \bibinfo{person}{Pearl Pu}.} \bibinfo{year}{2021}\natexlab{}.
\newblock \bibinfo{booktitle}{\emph{User Expectations of Conversational Chatbots Based on Online Reviews}}.
\newblock \bibinfo{publisher}{Association for Computing Machinery}, \bibinfo{address}{New York, NY, USA}, \bibinfo{pages}{1481–1491}.
\newblock


\bibitem[Tran et~al\mbox{.}(2019)]%
        {tran2019modeling}
\bibfield{author}{\bibinfo{person}{Jonathan~A. Tran}, \bibinfo{person}{Katie~S. Yang}, \bibinfo{person}{Katie Davis}, {and} \bibinfo{person}{Alexis Hiniker}.} \bibinfo{year}{2019}\natexlab{}.
\newblock \showarticletitle{Modeling the Engagement-Disengagement Cycle of Compulsive Phone Use}. In \bibinfo{booktitle}{\emph{Proceedings of the 2019 CHI Conference on Human Factors in Computing Systems}} (Glasgow, Scotland Uk) \emph{(\bibinfo{series}{CHI '19})}. \bibinfo{publisher}{Association for Computing Machinery}, \bibinfo{address}{New York, NY, USA}, \bibinfo{pages}{1–14}.
\newblock
\showISBNx{9781450359702}
\urldef\tempurl%
\url{https://doi.org/10.1145/3290605.3300542}
\showDOI{\tempurl}


\bibitem[Tversky and Kahneman(1974)]%
        {amos}
\bibfield{author}{\bibinfo{person}{Amos Tversky} {and} \bibinfo{person}{Daniel Kahneman}.} \bibinfo{year}{1974}\natexlab{}.
\newblock \showarticletitle{Judgment under Uncertainty: Heuristics and Biases}.
\newblock \bibinfo{journal}{\emph{Science}} \bibinfo{volume}{185}, \bibinfo{number}{4157} (\bibinfo{year}{1974}), \bibinfo{pages}{1124--1131}.
\newblock
\urldef\tempurl%
\url{https://doi.org/10.1126/science.185.4157.1124}
\showDOI{\tempurl}
\showeprint{https://www.science.org/doi/pdf/10.1126/science.185.4157.1124}


\bibitem[Van~Berkel et~al\mbox{.}(2017)]%
        {van2017experience}
\bibfield{author}{\bibinfo{person}{Niels Van~Berkel}, \bibinfo{person}{Denzil Ferreira}, {and} \bibinfo{person}{Vassilis Kostakos}.} \bibinfo{year}{2017}\natexlab{}.
\newblock \showarticletitle{The experience sampling method on mobile devices}.
\newblock \bibinfo{journal}{\emph{ACM Computing Surveys (CSUR)}} \bibinfo{volume}{50}, \bibinfo{number}{6} (\bibinfo{year}{2017}), \bibinfo{pages}{1--40}.
\newblock


\bibitem[van Berkel et~al\mbox{.}(2021)]%
        {berk}
\bibfield{author}{\bibinfo{person}{Niels van Berkel}, \bibinfo{person}{Mikael~B. Skov}, {and} \bibinfo{person}{Jesper Kjeldskov}.} \bibinfo{year}{2021}\natexlab{}.
\newblock \showarticletitle{Human-AI Interaction: Intermittent, Continuous, and Proactive}.
\newblock \bibinfo{journal}{\emph{Interactions}} \bibinfo{volume}{28}, \bibinfo{number}{6} (\bibinfo{date}{nov} \bibinfo{year}{2021}), \bibinfo{pages}{67–71}.
\newblock
\showISSN{1072-5520}
\urldef\tempurl%
\url{https://doi.org/10.1145/3486941}
\showDOI{\tempurl}


\bibitem[Weidinger et~al\mbox{.}(2022)]%
        {taxonomy}
\bibfield{author}{\bibinfo{person}{Laura Weidinger}, \bibinfo{person}{Jonathan Uesato}, \bibinfo{person}{Maribeth Rauh}, \bibinfo{person}{Conor Griffin}, \bibinfo{person}{Po-Sen Huang}, \bibinfo{person}{John Mellor}, \bibinfo{person}{Amelia Glaese}, \bibinfo{person}{Myra Cheng}, \bibinfo{person}{Borja Balle}, \bibinfo{person}{Atoosa Kasirzadeh}, \bibinfo{person}{Courtney Biles}, \bibinfo{person}{Sasha Brown}, \bibinfo{person}{Zac Kenton}, \bibinfo{person}{Will Hawkins}, \bibinfo{person}{Tom Stepleton}, \bibinfo{person}{Abeba Birhane}, \bibinfo{person}{Lisa~Anne Hendricks}, \bibinfo{person}{Laura Rimell}, \bibinfo{person}{William Isaac}, \bibinfo{person}{Julia Haas}, \bibinfo{person}{Sean Legassick}, \bibinfo{person}{Geoffrey Irving}, {and} \bibinfo{person}{Iason Gabriel}.} \bibinfo{year}{2022}\natexlab{}.
\newblock \showarticletitle{Taxonomy of Risks Posed by Language Models}. In \bibinfo{booktitle}{\emph{Proceedings of the 2022 ACM Conference on Fairness, Accountability, and Transparency}} (Seoul, Republic of Korea) \emph{(\bibinfo{series}{FAccT '22})}. \bibinfo{publisher}{Association for Computing Machinery}, \bibinfo{address}{New York, NY, USA}, \bibinfo{pages}{214–229}.
\newblock
\showISBNx{9781450393522}
\urldef\tempurl%
\url{https://doi.org/10.1145/3531146.3533088}
\showDOI{\tempurl}


\bibitem[Zeng et~al\mbox{.}(2021)]%
        {pol}
\bibfield{author}{\bibinfo{person}{Eric Zeng}, \bibinfo{person}{Miranda Wei}, \bibinfo{person}{Theo Gregersen}, \bibinfo{person}{Tadayoshi Kohno}, {and} \bibinfo{person}{Franziska Roesner}.} \bibinfo{year}{2021}\natexlab{}.
\newblock \showarticletitle{Polls, Clickbait, and Commemorative \$2 Bills: Problematic Political Advertising on News and Media Websites around the 2020 U.S. Elections}. In \bibinfo{booktitle}{\emph{Proceedings of the 21st ACM Internet Measurement Conference}} (Virtual Event) \emph{(\bibinfo{series}{IMC '21})}. \bibinfo{publisher}{Association for Computing Machinery}, \bibinfo{address}{New York, NY, USA}, \bibinfo{pages}{507–525}.
\newblock
\showISBNx{9781450391290}
\urldef\tempurl%
\url{https://doi.org/10.1145/3487552.3487850}
\showDOI{\tempurl}


\end{thebibliography}

\end{document}